\documentclass[12pt,english]{article}
\usepackage[english]{babel}
\usepackage{amsmath,amssymb,amscd,color,cite}
\usepackage{amsfonts}
\usepackage{graphicx}
\usepackage{url}
\usepackage{hyperref}

\makeatletter
\renewcommand\section{\@startsection {section}{1}{\z@}%
                                   {-5.5ex \@plus -1ex \@minus -.2ex}
                                   {2.3ex \@plus.2ex}%
                                   {\normalfont\large\bfseries}}
\renewcommand\subsection{\@startsection{subsection}{2}{\z@}%
                                     {-3.25ex\@plus -1ex \@minus -.2ex}%
                                     {1.5ex \@plus .2ex}%
                                     {\normalfont\bfseries}}

\numberwithin{equation}{section}

\makeatother

\newcommand{\bea}{\begin{eqnarray}}
\newcommand{\eea}{\end{eqnarray}}
\newcommand{\be}{\begin{equation}}
\newcommand{\ee}{\end{equation}}

\newcommand{\Z}{{\mathbb Z}}

\def\eg{{\it e.g.~}}

\newcommand{\no}[1]{\!:\! #1\!\! :}

\newcommand{\cC}{{\cal C }}
                
\newcommand{\cO}{{\cal O }}

\newcommand{\cG}{{\cal G }}
\newcommand{\cN}{{\cal N }}

\newcommand{\cI}{{\cal I }}

\newcommand\nn{\nonumber}
\def\({\left(} 
\def\){\right)}

\newcommand{\vac}{|0\rangle}

\newcommand{\wb}{{\bar w}}
\newcommand{\zb}{{\bar z}}
\newcommand{\xb}{{\bar x}}
\newcommand{\hb}{{\bar h}}

\newcommand{\pl}[2][]{\Psi^{(#2)#1}}
\newcommand{\pbl}[2][]{\bar{\Psi}^{(#2)#1}}
\newcommand{\pr}[2][]{\tilde{\Psi}^{(#2)#1}}
\newcommand{\pbr}[2][]{\tilde{\bar{\Psi}}^{(#2)#1}}

\newcommand{\Xl}[2][]{\partial X^{(#2)#1}}
\newcommand{\Xbl}[2][]{\partial\bar{X}^{(#2)#1}}
\newcommand{\Xr}[2][]{\partial\tilde{X}^{(#2)#1}}
\newcommand{\Xbr}[2][]{\partial\tilde{\bar{X}}^{(#2)#1}}

\newcommand{\notebook}{\texttt{lifting.nb}}

\begin{document}
	
	\begin{titlepage}
		\begin{center}
			
			\hfill \\
			\hfill \\
			\vskip 0.75in
			
			{\Large 
				\bf Lifting 1/4-BPS States in $AdS_3\times S^3 \times T^4$
			}\\

			\vskip 0.7in
			
			{\large Nathan Benjamin,$^{a}$ Christoph A.~Keller,${}^{b}$  and Ida G.~Zadeh${}^{c}$
			}\\
			\vskip 0.4in
			
			${}^a$
			{\it Princeton Center for Theoretical Science, Princeton University, 
				Princeton, NJ 08544, USA}\vskip 1mm
			${}^{b}$
			{\it Department of Mathematics, University of Arizona, Tucson, AZ 85721-0089, USA} \vskip 1mm
			${}^{c}$
			{\it International Centre for Theoretical Physics, Strada Costiera 11, 34151 Trieste, Italy} 
			
			\vskip 0.2in

			\texttt{nathanb@princeton.edu, cakeller@math.arizona.edu, zadeh@ictp.it}

		\end{center}
		
		\vskip 0.35in
		
		\begin{center} {\bf ABSTRACT } 	\end{center}
We establish a framework for doing second order conformal perturbation theory for the symmetric orbifold Sym$^N(T^4)$
to all orders in $N$. This allows us to compute how 1/4-BPS states of the D1-D5 system on $AdS_3\times S^3\times T^4$ are lifted as we move away from the orbifold point. As an application we confirm a previous observation that in the large $N$ limit not all 1/4-BPS states that can be lifted do get lifted. This provides evidence that the supersymmetric index actually undercounts the number of 1/4-BPS states at a generic point in the moduli space.

		\vfill
		
		\noindent July 12, 2021

	\end{titlepage}
	\tableofcontents

\section{Introduction and Summary of Results}

The D1-D5 system is one of the best studied examples of the AdS/CFT correspondence \cite{Maldacena:1997re}. Here string theory on an AdS$_3 \times S^3 \times X$ background is dual to the symmetric product orbifold of $X$ for $X = T^4$ or K3. However, at the symmetric orbifold locus itself, the theory is very stringy: it is dual to a tensionless string theory rather than an Einstein gravity theory (see \cite{Eberhardt:2018ouy,Eberhardt:2019ywk,Eberhardt:2020bgq} for an exact duality between the symmetric orbifold and a tensionless string with one unit of NS-NS flux when $X=T^4$). In fact, any symmetric orbifold theory has a low-lying density of states growing exponentially with energy rather than sub-exponentially, which is consistent with a string theory rather than supergravity interpretation \cite{Keller:2011xi, Hartman:2014oaa}. However, it is believed that deforming the symmetric product orbifold theory by an exactly marginal operator (modulus) in the twisted sector of the orbifold takes the theory to a strongly-coupled regime in which most of the low-lying states get lifted, making the spectrum compatible with a supergravity interpretation. 

In view of this, the purpose of this article is twofold. First, we establish a systematic framework for second order perturbation theory of the symmetric product orbifold of $T^4$. Such computations have of course been done before, {\it e.g.} in \cite{Dijkgraaf:1987jt,Cardy:1987vr,Kutasov:1988xb,Eberle:2001jq,Gava:2002xb,Pakman:2009mi}, but our framework allows to do them systematically to all order in $N$. The crucial ingredient here is a careful analysis of the combinatorial factors that appear due to the symmetric orbifold. To our knowledge, this is the first time this has been done in the literature. 
In the process we focus on 1/4-BPS primary fields and their lifting. Our methods are particularly appropriate for this for two reasons: First, because 1/4-BPS states saturate the unitarity bound, the first order perturbation term has to vanish, so that the second order term is the leading contribution. Second, we can use Ward identities to simplify the problem considerably. In particular, the area integral of 4-point functions which have to be evaluated at the second order can be reduced to a non-holomorphic contour integral, which then only picks out a finite number of terms. This leads to a slightly more general version of the Gava-Narain formula for the lifting matrix $D^{k\ell}$ first found in \cite{Gava:2002xb} and also used in \eg \cite{deBoer:2008ss,Guo:2019pzk},
\be
D^{k\ell}
=\sum_{\substack{\chi:h_\chi=h_k\\ \hb_\chi=\hb_k+1/2}}
2\pi C^*_{\chi^\dagger O^{'}\varphi^{\ell}} C_{\chi^\dagger O'\varphi^k}
+\sum_{\substack{\chi:h_\chi=h_k\\ \hb_\chi=\hb_k-1/2}}
2\pi C^*_{\chi^\dagger  O^{'\dagger}\varphi^{\ell}} C_{\chi^\dagger O^{'\dagger}\varphi^k}\ .
\ee
We will give a detailed description of the ingredients in section~\ref{s:2ndorder}, but the basic idea is that the lifting matrix $D^{k\ell}$ can be written as a finite sum of squares of 3-point functions involving the lifted fields $\varphi^k$ and the moduli $O$. Depending on the situation this formula may or may not be more efficient than computing the full 4-point function, but it is certainly a useful way of exhibiting structural properties such as positive definiteness of the lifting matrix. 

We explain how to perform such computations for symmetric product orbifolds of $T^4$, including the full combinatorial factors that give the all order $N$ dependence, and also explain how to compute the necessary 4-point functions in practice by using the well-known trick of going to a cover surface. In a few simple cases we go through the computations in detail. For more complicated cases, we provide a Mathematica notebook, \notebook, which can perform such computations in great generality.

The second purpose of this article is an application of the framework we established. The goal here is to understand the lifting of 1/4-BPS states in the moduli space of Sym$^N(T^4)$.
For this, consider the elliptic genus, which is an index of 1/4-BPS states. The elliptic genus of Sym$^N(K3)$ can be computed \cite{Dijkgraaf:1996xw}, and for large $N$ has the asymptotic behavior
\be
\rho(\Delta) \sim e^{\#\sqrt{\Delta}}, ~~~~~~ 1 \ll \Delta \ll N\ .
\label{eq:EGScaling}
\ee
The elliptic genus however does not count 1/4-BPS, but rather computes an index, which potentially has a lot of cancellations. The number of 1/4-BPS states of Sym$^N(K3)$ in fact grows much faster \cite{Benjamin:2016pil} , 
\be
\rho(\Delta) \sim e^{2\pi \Delta}, ~~~~~~ 1 \ll \Delta \ll N\ .
\label{eq:CKScaling}
\ee
This much faster Hagedorn growth is actually the same as for non-BPS states \cite{Keller:2011xi}. There is an immediate physical argument why this faster growth is not allowed for the elliptic genus: since the elliptic genus is a protected quantity and therefore constant on the entire moduli space, if it did grow as (\ref{eq:CKScaling}), then there could never be a point in its moduli space where the theory is dual to large-radius Einstein gravity \cite{Benjamin:2015hsa}.  Interestingly this is a somewhat unusual property of K3. For most supersymmetric ``seed" CFTs $X$, the growth of the low-lying states of both the partition function and the elliptic genus of Sym$^N(X)$ has the same scaling behavior \cite{Benjamin:2015vkc, Belin:2019rba, Belin:2019jqz, Belin:2020nmp}. 

In view of this a natural question to ask is: at a generic point in moduli space, is the growth of BPS sates as in (\ref{eq:EGScaling}), or is it somewhere between (\ref{eq:CKScaling}) and (\ref{eq:EGScaling})? To address this question it is instructive to consider the supergravity point. If instead of computing the index, we simply count the number of supergravity KK modes in AdS$_3 \times S^3$ \cite{deBoer:1998kjm, deBoer:1998us}, the growth of BPS states scales as \cite{Benjamin:2016pil}
\be
\rho(\Delta) \sim e^{\# \Delta^{3/4}}, ~~~~~~ 1 \ll \Delta \ll N\ ,
\label{eq:BPSScaling}
\ee
which although slower than (\ref{eq:CKScaling}), is still parametrically faster than (\ref{eq:EGScaling}). If the same thing holds not just at the supergravity point but also at a generic point in the moduli space, then this would imply that many BPS states that \emph{can} cancel from the representation theory of the $\mathcal{N}=4$ superconformal algebra (SCA) nevertheless do \emph{not} cancel even at a generic point. 

The question of whether or not a protected signed count of supersymmetric states is representative of the true density of states is of course of great interest to many physically important problems including black hole entropy \cite{Strominger:1996sh}, and recently has had renewed interest \cite{Kinney:2005ej, Choi:2018hmj}. 
We want to address this question by conformal perturbation theory for the case $X=T^4$. Here (\ref{eq:CKScaling}) and (\ref{eq:BPSScaling}) also hold, and we want to investigate for points near but not on the orbifold locus which one holds. For this we want to investigate how many of the 1/4-BPS states get lifted as we deform the theory away from the symmetric orbifold point to a more generic point.

Let us point out that this type of analysis can be done for simpler examples, such as for K3 and its description as a $T^4/\Z_2$ orbifold. The elliptic genus of K3 predicts for instance that there are 90 1/4-BPS states of weight $h=1$. K3 is  special in the sense that there only two short representations, one of which is the vacuum. Cancellations that occur must therefore involve chiral fields, that is currents corresponding to additional symmetries. This is for instance what happens at the $T^4/\Z_2$ orbifold locus of K3: here there are additional chiral fields which lead to 102 rather than 90 $h=1$ 1/4-BPS states. Assuming that there are no such chiral fields at a generic point, \cite{Ooguri:1989fd} computed the 1/4-BPS spectrum at a generic point in the moduli space. This prediction was recently confirmed in \cite{Keller:2019suk}: by perturbing away from the orbifold point, it was established that 12 of the 102 states pair up to non-BPS states and get lifted, so that only the predicted 90 states remain BPS. We note that this question recently became important in the context of Mathieu moonshine \cite{Eguchi:2010ej} and its proposed explanation through symmetry surfing \cite{Taormina:2013jza,Taormina:2013mda}.

In this article we want to consider Sym$^N(T^4)$. Here the situation is a bit different from the K3 example just discussed. First of all, the elliptic genus vanishes; instead we could compute the ``modified" index of \cite{Maldacena:1999bp}. The elliptic genus of Sym$^N(T^4)$ vanishes due to fermionic zero-modes. Relatedly, the theory has not just small $\cN=4$ superconformal symmetry, but actually an enlarged chiral algebra called the contracted large $\cN=4$ superconformal algebra. Using this fact, we can be more precise about the lifting of 1/4-BPS states: short representations $\chi_j$ of the contracted large $\cN=4$ SCA can combine to long representations $\chi_{h,j}$ as (see section \ref{ss:N4} for notations)
\be\label{shorttolong}
\chi_j + 2\chi_{j+1} + \chi_{j+2}=\chi_{j/2,j}\ .
\ee
The modified index of \cite{Maldacena:1999bp} mentioned above is indeed invariant under (\ref{shorttolong}).

In this article we focus on 1/4-BPS states in the $h=1,j=0$ long representation for the left movers, and in the $\bar j$ short representation for the right movers. At the symmetric product orbifold point the number of primaries is listed in the first line of table~\ref{t:14BPSintro} As expected, there are many such states at the symmetric orbifold point. Generically, one might expect that when perturbing away from that point, all short representations would combine to long representations according to (\ref{shorttolong}) whenever possible. Because there are now more than just two representations, the situation is now more complicated than in the K3 example above. In particular just assuming that there are no chiral fields at a generic point is no longer enough to fix the entire 1/4-BPS spectrum. Nonetheless one could conjecture that as many states as possible get lifted. This would lead to a `minimal' spectrum as given in the second line of table~\ref{t:14BPSintro}. In computing this minimal spectrum, we assume that there are no additional conserved currents in the spectrum, and that we are only allowed to pair states up in accordance with (\ref{shorttolong}) from the orbifold point; the minimal spectrum is not necessarily unique.
\begin{table}[ht]
	\centering
	\begin{tabular}{|c|ccccccc|}
		\hline
		$\bar j$&0&1&2&3&4&5& $\geq 6$ \\
		\hline
		symmetric orbifold &3&14&36&44&26&6&0\\
		minimal&0&0&3&2&0&0&0\\
		sugra&0& 0& 10& 20& 15& 4& 0\\
		\hline
	\end{tabular}
	\caption{1/4-BPS spectrum for $h=1, j=0$ left-moving states.}\label{t:14BPSintro}
\end{table}
However, in view of the remarks above, this expectation is too quick. 
\cite{Benjamin:2017rnd} also computed the 1/4-BPS spectrum at the supergravity point. The result is given in the third line of table~\ref{t:14BPSintro}, and we immediately see that more than just the minimal number of states remain unlifted.
For example additional 10 cancellations of the form
\be
\chi_{1,0} \tilde{\chi}_{2} + 2\chi_{1,0} \tilde{\chi}_{3} + \chi_{1,0} \tilde{\chi}_{4} = \chi_{1,0} \tilde{\chi}_{1, 2}
\ee
in principle could occur; here $\tilde \chi$ denotes the right-moving characters. Given that these cancellations do not occur at the supergravity point, it is reasonable to conjecture that not all 1/4-BPS states get lifted under perturbation away from the orbifold point. 

This is indeed what we will check in our paper. For simplicity we check only states in the untwisted sector at the orbifold point. A more refined count separating the BPS states at the symmetric product orbifold point into untwisted vs. twisted sectors is given in table \ref{t:14BPS} in section~\ref{s:liftUt}. 
At leading order in $1/N$, we find that all 3 states with $\bar j=0$ and all 6 untwisted states with $\bar j=1$ do indeed get lifted. However, we find that for $\bar j =2$, only 3 of the 9 untwisted 1/4-BPS states are lifted at second order in perturbation theory. This means that at large $N$, the total number of 1/4-BPS states cannot be the minimal spectrum as in table \ref{t:14BPSintro}, but it is compatible with the result at the supergravity point.


Let us be slightly more careful here. The reason why we talk about leading order in $N$ is because we want to compare to the supergravity result, which requires taking the large $N$ limit. More precisely, the correct prescription is to keep the 't Hooft type coupling $\lambda/\sqrt{N}$ fixed while sending $N\to\infty$; here $\lambda$ is the coupling that appears in the conformal perturbation theory. At second order perturbation theory this means that we are keeping only the $N^{-1}$ term and discard all subleading terms.

When we do this, as described above at $\bar j=2$ indeed only 3 of the 9 untwisted states get lifted. If however we keep all orders in $N$, 6 of the 9 untwisted sector states are lifted. If every single twisted sector state gets lifted, then this would still be consistent with the minimal spectrum in table \ref{t:14BPSintro}. It is thus possible that for finite $N$, generically all possible liftings occur. We note that this agrees with the results of \cite{Guo:2019ady}, which did the computation for $N=2$ and found liftings compatible with the minimal spectrum --- see appendix~\ref{app:quartBPS} for a discussion of this. To fully verify this, we need to compute the lifting of the twisted sector states, which we leave to future work.
We also note that there is a nice supergravity interpretation of this: The weight of multi particle states is protected to leading order in $N$, but interactions between protected single particle states will lead to corrections at subleading order, which is exactly what we find here.


\subsection{Comparison to Literature}\label{ss:lit}
Before starting out, let us briefly discuss how our results fit in with the literature on conformal perturbation theory for holographic CFTs. In particular we consider the following works:

\begin{itemize}
	\item \cite{Gaberdiel:2015uca} computed the lifting of single trace chiral states, that is higher spin fields with $h=s$ and $\hb=0$. They restricted to states which are uncharged with respect to the large $\mathcal N=4$ SCA. In particular, they computed the lifting of the `flavor' ${su}(2)$ symmetry current $\hat J^{(3)}$ (see eq. (\ref{Jhatminus}) below), and find its lifting:
	\be
	\delta h = \frac12 \frac{\lambda^2\pi^2}{N}\ .
	\ee
	(Note that in their notation they replace $N$ by $N+1$.)
	They only consider the single trace part of the state, and our result agrees with their leading order result up to a factor of 2, which we believe is related to the choice of the normalisation of the twisted sector operators --- see section \ref{subsec_norm} below.

	\item \cite{Guo:2019pzk,Guo:2019ady,Guo:2020gxm} developed technology and computed lifting of states in the Ramond sector which are in the right moving Ramond ground state, that is have conformal dimensions $(h,c/24)$. In \cite{Guo:2019ady} states with $h=1$ and $N=2$ copies of the seed theory are considered. They found that a triplet of long multiplets get lifted by the same amount
	\be
	\delta E = \lambda^2\pi^2\ .
	\ee
The computation was extended to $h=4$ in \cite{Guo:2020gxm} and it was found that, up to second order in perturbation theory, all the states that can lift (by the supersymmetric index) indeed get lifted. We perform our computations in the NS sector, but of course the results are directly related by spectral flow. The triplet they consider corresponds again to our flavor currents $\hat J$, and we indeed also find that they all get lifted by an equal amount. We did however not match the amount of lifting to their their results exactly, since our result strictly speaking only holds for $N\geq 3$. In fact the case $N=2$ is special enough that under a weak assumption one can compute the 1/4-BPS spectrum at a generic point directly. We discuss this in appendix~\ref{app:quartBPS}.

	\item \cite{Hampton:2018ygz} computed lifting of particular single particle and multi particle states in the untwisted sector which have excitations of the form $J^{(+)}_{-(2m+1)}\cdots J^{(+)}_{-3}J^{(+)}_{-1}|0\rangle_{\text NS}$ on the NS vacuum of various copies of the seed theory.

	\item In a series of papers \cite{Lima:2020boh,Lima:2020nnx,Lima:2020kek,Lima:2020urq,Lima:2021wrz} lifting of various twisted sector Ramond fields and some of their composite operators are computed up to second order in perturbation theory.
\end{itemize}

The organization of the rest of the paper is as follows: In section \ref{sec_sym} we define our notation for the symmetric product orbifold CFT, the ${\cal N}=4$ SCA, as well as the normalization of the states. In section \ref{s:2ndorder}, we discuss our setup for second order conformal perturbation theory, the superconformal Ward identities we use, and the evaluation of the contour integrals. Section \ref{s:corrcomp} describes the computation of the 4-point functions of the symmetric product orbifold CFT. Finally, in section \ref{s:liftUt} we compute lifting of a set of 1/4-BPS states in the untwisted sector with $h=1$ and $j=0$, and with $\bar j=0, 1,2$. In appendix \ref{app:quartBPS} we discuss the representation theory of the small and contracted large ${\mathcal N}=4$ SCA and in appendix \ref{app:3pt} we compute 3-point functions involved in the computation of the lifting of the 3 currents with $h=1$ and $j=\bar j=0$. An ancillary Mathematica notebook, \notebook, performs the computation of lifting at the second order in perturbation theory.

\section{The Symmetric Orbifold of $T^4$: Setup and Notation}\label{sec_sym}

\subsection{Conventions and Notation}
Unfortunately symmetric orbifold computations tend to be swamped by the number of different indices. We are taking the following conventions for fields $\phi$: Downstairs indices $\phi_n$ always denote the modes. Upstairs indices without round parentheses $\phi^i$ always denote the tensor factor, so that $\phi=\bigotimes_{i=1}^N \phi^i$. The symmetric group $S_N$ acts by permuting the tensor factors, 
\be
g\phi := \bigotimes_{i=1}^N \phi^{g(i)}\ ,
\ee
potentially introducing fermionic signs in the process.
Operators in the symmetric orbifold are given by orbits under this action.
We will denote by $\check \phi$ a representative of that orbit, usually chosen such that the all non-vacuum factors are in the front. The actual symmetrized operator $\phi$,  up to normalization, is then given by 
\be
\phi \sim \sum_{g\in S_N} g\check \phi\ .
\ee
More precisely, pick a representative of this orbit $\check \phi$ whose first $L$ factors are non-vacuum factors, and whose last $N-L$ factors are vacua,
\be
\check\phi = \phi^1\otimes \cdots \phi^L\otimes \bigotimes^{N-L}\vac\ .
\ee
Note we can use $\check \phi$ to construct such a state for any $N$, as long as $N\geq L$. This allows to construct a large $N$ limit.
A normalized state $\phi$ which is permutation symmetric can then be written as the normalized orbit of $\check \phi$,
\be\label{orbitstate}
\phi = \frac1{\sqrt{N! A_{\phi}(N-L)!}}\sum_{g\in S_N} g\check\phi
\ee
where the normalization constant $A_\phi$ is independent of $N$, and is given by
\be
\sum_{g\in S_L}\langle g\check\phi|\check\phi\rangle = A_{\phi}
\ee
Upstairs indices in round parentheses $\phi^{(i)}$ are additional labels, for instance the 4 coordinates $I=1,2,3,4$ of the $T^4$. Right-movers are denoted by a tilde, $\tilde \phi$. Bar denotes the complex conjugation with respect to the complex coordinates on $T^4$. 

Finally, let us introduce the convention that we call states with $L=1$, that is with only one non-vacuum factor, \emph{single trace}, and states with $L>1$ \emph{multi-trace}. The rationale behind this name is that in the large $N$ limit, the correlation functions of such states behave exactly like single trace and multi-trace operators in large $N$ SYM: to leading order, correlators of multi-trace operators can be computed as Wick contractions of their single trace components \cite{Belin:2015hwa}.

\subsection{The Contracted Large $\cN=(4,4)$ SCA for $T^4$ and Sym$^N(T^4)$}\label{ss:N4}
Next let us introduce the contracted large $\cN=(4,4)$ superconformal algebra and its realization on $T^4$ and Sym$^N(T^4)$.
The contracted large $\cN=4$ SCA has the generators 
\be\label{xpsiI}
\Xl{I}\ , \ \pl{I}\ , \ L\ , \ G^{(\alpha A)} \ ,\ J^{(a)}\ ,
\ee
where $I=1,2,3,4$, $\alpha=\pm$, $A=1,2$, and $a=1,2,3$ or $a=\pm,3$.
Its central charge is $c=6N$.

Let us first discuss its structure for $c=6$. In this case the $\Xl{I}$ and $\pl{I}$ are simply given by four free fermions and four free bosons on ${T}^4$. We will write them in complex notation
\be
\pl{i} \ , \quad \pbl{i}  \ , \qquad  \Xl{i}\ , \quad \Xbl{i}  \ ,
\ee
where $i=1,2$ with the understanding that $\pl{3}:=\pbl{1}$ and $\pl{4}:=\pbl{2}$ and likewise for the bosons. The non-vanishing (anti-)commutation relations are given by
\be\label{Xcommu}
[\Xl{i}_m,\Xbl{j}_n]=m\delta^{ij}\delta_{m,-n}
\ee
\be\label{psianticommu}
\{\pl{i}_r,\pbl{j}_s \}=\delta^{ij}\delta_{r,-s}
\ee
The Virasoro tensor $L$ is given by the usual expression.
The R-symmetry ${su}(2)$ currents are given by
\begin{eqnarray}\label{J}
J^{(3)} & = & \frac{1}{2} \Bigl(  :\pl{1} \pbl{1}: + :\pl{2} \pbl{2}:\Bigr) = \frac{1}{2} \delta_{ij}:\pl{i}\pbl{j}:\ ,\\[2pt]
J^{(+)} & = & - \pl{1} \pl{2} = -\frac{1}{2}\epsilon_{ij}\pl{i}\pl{j}\nonumber\ ,\\[2pt]
J^{(-)} & = & \pbl{1} \pbl{2} =\frac{1}{2}\epsilon_{ij}\pbl{i}\pbl{j}\ ,\nonumber
\end{eqnarray}
where we work in the usual Cartan-Weyl basis for ${su}(2)$.
Finally the four supercurrents $G^{(\alpha A)}$ are given by
\begin{eqnarray}\label{Gs}
G^{(+1)} = \delta_{ij}\pl{i}\Xbl{j} && G^{(-1)} = \epsilon_{ij}\pbl{i}\Xbl{j}\\
G^{(+2)} = -\epsilon_{ij}\pl{i} \Xl{j} && G^{(-2)} = \delta_{ij}\pbl{i}\Xl{j}\nonumber
\end{eqnarray}
The index $\alpha=\pm$ indicates the R-charge, that is the $G^{aA}$ form two doublets under the ${{su}}(2)$  R-symmetry.
Note that there is also a second  `flavor' ${{su}}(2)$ symmetry which is an outer automorphism  of the small $\cN=4$ superconformal algebra. The corresponding currents $\hat J$ are given by
\begin{eqnarray}\label{Jhatminus}
\hat J^{(3)} & = & \frac{1}{2} \Bigl( - \pl{1} \pbl{1} + \pl{2} \pbl{2} \Bigr)\ ,\\[2pt]
\hat J^{(+)} & = & - \pl{1} \pbl{2} \ ,\nonumber \\[2pt]
\hat J^{(-)} & = & \pbl{1} \pl{2} \ .\nonumber
\end{eqnarray}

For $c>6$, we can describe the algebra in a way that is very natural for the symmetric orbifold. Namely, we can write the generators of the $c=6N$ $\cN=4$ algebra as single trace version of the generators of the $c=6$ algebra, namely \be\label{symSum}
O = \sum_{i=1}^N O^i\ .
\ee
Note that in the case $c=6$ the generators $L,G,J$ could be expressed in terms of the free fields $\Xl{I}$ and $\pl{I}$.
For $c>6$ this is no longer the case: now $L,G,J$ can no longer be written in terms of the single trace version of $\Xl{I}$ and $\pl{I}$, since their bilinears are multi-trace states. This means that the single trace versions of $L,G,J$ are new, independent generators.
\cite{Hampton:2018ygz,Guo:2019ady}

Note that the $\cN=4$ SCA of course still contains the R-symmetry $su(2)$ currents $J$. These are expected to survive under perturbations away from the orbifold point; in fact we check this explicitly at second order in section~\ref{ss:j0lift}. The situation for the flavor symmetry currents $\hat J$ however is different: They are still present at the orbifold point, where they are given by the single trace version of the seed generators $\hat J$. They are however \emph{not} part of the $\cN=4$ SCA, and are therefore not expected to survive perturbations away from the orbifold point. In fact, this is exactly what the result of \cite{Guo:2019ady} implies for $N=2$, and what we will confirm in section~\ref{ss:j0lift}: the generators of $\hat J$ get lifted.

Let us finally discuss the representation theory of the contracted large $\cN=4$ SCA.
We will perform all our computations in the NS sector. In the NS sector, the large $\cN=4$ SCA has a
family of irreducible short representations 
\be
\chi_j\qquad j=0,1,\ldots c/6-1\ ,
\ee
which have $su(2)$ spin $j$ and conformal weight $h=j/2$,
and a family of irreducible long representations
\be
\chi_{h,j}\qquad j=0,\ldots c/6-2 \textrm{ with }\ h>j/2\ , 
\ee
with $su(2)$ spin $j$ and conformal weight $h$ \cite{Petersen:1989zz, Petersen:1989pp}.
As usual, at the unitarity bound a long representation decomposes into short representations,
\be
\chi_j + 2\chi_{j+1} + \chi_{j+2}=\chi_{j/2,j}\ .
\ee
The conformal weight of short representations is of course fixed, so that they will not be lifted. Once they combine into long representations, their weight is no longer protected, and they can be lifted. This is exactly what we are probing with perturbation theory.

\subsection{Twist 2 sector}\label{ss:twist2}
Let us set up the computation for lifting untwisted states for $T^4$. First, let us discuss the modulus $O$. 
The $N^{\text{th}}$ symmetric orbifold has twist 2 sectors which are denoted by permutations $(ij)$ of cycle length 2. Each such sector has twisted ground state $\sigma_{(ij)}$. For concreteness let us first discuss the twist field $\sigma_{(12)}$. It acts by permuting the first and second tensor factor. That is, when rotating an operator $\phi^1$ around it, the operator gets mapped to $\phi^2$ and vice versa. To make the connection to the usual $\Z_2$ orbifold, it is useful introduce a change of basis for fields $\phi^{1,2}$,
\be\label{phiSA}
\phi^S = \frac1{\sqrt{2}}(\phi^1+\phi^2)
\qquad \phi^A = \frac1{\sqrt{2}}(\phi^1-\phi^2)
\ee
$\phi^S$ is invariant under rotations around $\sigma_{(12)}$, whereas $\phi^A$ picks up a minus sign. In particular this implies that its moding will be different. 

The twist field $\sigma_{(12)}=\sigma_b\sigma_f$ has a bosonic $\sigma_b$ and a fermionic component $\sigma_f$. The bosonic component is uncharged and has the usual dimension
\be
h_{\sigma_b}=\hb_{\sigma_b} = \frac c{24}(n-1/n)= 1/4\ .
\ee
To describe the fermionic component, it is useful to bosonize the fermions $\pl[A]{i}$.
We define
\bea\label{bsnsn_2compferms}
&&\pl[A]{1}=e^{iH^{(1)}}\ ,\qquad\pbl[A]{1}=e^{-iH^{(1)}}\ ,\\
&&\pl[A]{2}=e^{iH^{(2)}}\ ,\qquad\pbl[A]{2}=e^{-iH^{(2)}}\ ,\nonumber
\eea
where $H^{(1)}$ and $H^{(2)}$ are real bosonic fields. We actually need to be slightly careful here to make sure that two orthogonal fermionic operators anti-commute rather than commute. One way to deal with this issue is to introduce anti-commuting Klein factors $\eta_i$ satisfying $\{\eta_i,\eta_j\}=2\delta_{ij}$ \cite{Senechal:1999us}.
Next note that $\Psi^A$ is anti-periodic and therefore has zero modes. This means that $\sigma_f$ actually is a 16-dimensional space on which the zero modes $\pl[A]{i}_0,\pbl[A]{i}_0, i=1,2$ and their right-moving counterparts act. In terms of the bosonized momenta the states in this space are given by
\be
|k_1,k_2;\tilde k_1,\tilde k_2\rangle \qquad k_i, \tilde k_i\  =-\frac12,\frac12\ .
\ee
We take the convention that 
\be
\pl{1}_0\pl{2}_0|-\frac12,-\frac12;\tilde k_1,\tilde k_2\rangle
=|\frac12,\frac12;\tilde k_1,\tilde k_2\rangle
\ee
The corresponding operators are given by
\be\label{sigmaf}
\sigma_f^{k_1k_2\tilde k_1 \tilde k_2} = \eta_1^{k_1+1/2} e^{ik_1H^{(1)}} \eta_2^{k_2+1/2}e^{ik_2H^{(2)}} \eta_3^{\tilde k_1+1/2}e^{i\tilde k_1\tilde H^{(1)}} \eta_4^{\tilde k_2+1/2}e^{i\tilde k_2 \tilde H^{(2)}}\ , 
\ee
That is, we took the convention that operators with positive values of $k$ have odd fermion parity.
Let us call $\sigma^+_f = \sigma_f^{\frac12\frac12\frac12\frac12}$ and $\sigma^-_f = \sigma_f^{-\frac12-\frac12-\frac12-\frac12}$ its Hermitian conjugate $(\sigma_f^+)^\dagger= \sigma_f^-$. $\sigma^-_f$ is thus annihilated by all $\pbl[A]{i}_0,\pbr[A]{i}_0$.
From (\ref{sigmaf}) we immediately see that $h_{\sigma_f}=\hb_{\sigma_f} = 1/4$, so that $h_{\sigma_{(12)}}=\hb_{\sigma_{(12)}}=1/2$. Moreover $\sigma_{(12)}$ is an $su(2)$ doublet for both the left- and right-moving R-symmetry. It is thus a 1/2-BPS state, exactly as expected for a modulus.

\subsection{Normalization and Permutation Symmetry}\label{subsec_norm}

For the twist 2 ground state $\sigma_2$, we choose $\check \sigma_2 = \sigma_{(12)}$ and get
\be
\sigma_2 = \frac1{\sqrt{N!2(N-2)!}}\sum_{g\in S_N} \sigma_{(g(1)g(2))}\ .
\ee
To obtain the actual modulus $\cO$, we need to act on it with appropriate supercharges $G_{-1/2}$ and $\tilde G_{-1/2}$. In fact, the primary $\sigma_2$ leads to 4 possible moduli, which in $\cN=2$ language correspond to (c,c), (a,c), (c,a) and (a,a) chiral rings. For concreteness we will pick $\sigma=\sigma_b\sigma_f^-$, and act with $G^{(+1)}_{-1/2}$ as the $G$ descendant. We have
\begin{multline}\label{Omodulus}
G^{(+1)} = \sum_{i=1}^N\left( \pl[i]{1} \Xbl[i]{1}+\pl[i]{2} \Xbl[i]{2} \right)\\
=\pl[S]{1} \Xbl[S]{1}+\pl[S]{2} \Xbl[S]{2}+\pl[A]{1} \Xbl[A]{1}+\pl[A]{2} \Xbl[A]{2}+ \sum_{i=3}^N \left(\pl[i]{1} \Xbl[i]{1}+\pl[i]{2} \Xbl[i]{2}\right)
\end{multline}
We then have
\begin{multline}\label{Odaggermodulus}
\check O = G^{(+1)}_{-1/2}\tilde G^{(+1)}_{-1/2}\sigma_{(12)}^-
= \\ (\Xbl[A]{1}_{-\frac12}\pl[A]{1}_0 + \Xbl[A]{2}_{-\frac12}\pl[A]{2}_0)(\Xbr[A]{1}_{-\frac12}\pr[A]{1}_0 + \Xbr[A]{2}_{-\frac12}\pr[A]{2}_0)\sigma_{(12)}^-\ .
\end{multline}
Note that the symmetric contribution to $G_{-1/2}$ is of the form $\partial X^S_0 \Psi^S_{-1/2}$, and therefore annihilates $\sigma_{(12)}$, so that only the anti-symmetric contribution survives. The hermitian conjugate of the modulus is
\begin{multline}
\check O^\dagger =G^{(-2)}_{-1/2}\tilde G^{(-2)}_{-1/2}\sigma_{(12)}^+\\
= (\Xl[A]{2}_{-1/2}\pl[A]{1}_0 - \Xl[A]{1}_{-1/2}\pl[A]{2}_0)
(\Xr[A]{2}_{-1/2}\pr[A]{1}_0 - \Xr[A]{1}_{-1/2}\pr[A]{2}_0)\sigma_{(12)}^-\ .
\end{multline}
It is straightforward to check that $\check O$ is properly normalized, $\langle \check O|\check O\rangle$.
In total, the modulus is thus given by
\be\label{Onorm}
O = \frac1{\sqrt{N!2(N-2)!}}\sum_{g\in S_N} g\check O\ ,
\ee
and the hermitian linear combination of the two moduli are:
\be\label{ccmodulus_real}
\cO\equiv\textstyle\frac{1}{\sqrt{2}}(O+O^\dagger)\ , \qquad \widehat{\cO}\equiv\frac{i}{\sqrt{2}}(O-O^\dagger)\ .
\ee

\section{Second Order Perturbation Theory for 1/4-BPS States}
\label{s:2ndorder}
\subsection{Setup}
Let us first set up our notation for computing the shift in weight of 1/4-BPS states in a general CFT.
In setting up our conventions for conformal perturbation theory, we mostly follow \cite{Keller:2019yrr}.
We perturb the action by
\be
S + \lambda \int d^2z \cO(z,\zb)\ ,
\ee
where the normalized modulus $\cO$ is given by one of the hermitian linear combinations in (\ref{ccmodulus_real}).
For concreteness we will work with the combination $\cO$.
We want to compute how the weight of set of 1/4-BPS states $\varphi^{(k)}$ gets shifted at second order. Because these will in general be degenerate, we need to take into account operator mixing.
We therefore introduce the lifting matrix
\be\label{gamma}
\gamma^{k\ell} = \frac\pi4 \lambda^2 D^{k\ell}
\ee
where we define the matrix
\be\label{DijGij}
D^{k\ell}:=-\int d^2 x \mathcal G^{k\ell}(x,\xb) = -\frac i2 \int dx d\xb \cG^{k\ell}(x,\xb)
\ee
where $\cG^{k\ell}(x,\xb)=\cG^{k\ell}_1(x,\xb)+\cG_2^{k\ell}(x,\xb)$ with
\bea
\cG^{k\ell}_1(x,\xb) &=& \langle \varphi^{(\ell)\dagger}(\infty,\infty)O^\dagger(1,1)O(x,\xb)\varphi^{(k)}(0,0)\rangle\ ,\label{curlg1}\\ 
\cG^{k\ell}_2(x,\xb) &=& \langle \varphi^{(\ell)\dagger}(\infty,\infty)O(1,1)O^\dagger(x,\xb)\varphi^{(k)}(0,0)\rangle\ .\label{curlg2}
\eea
The lifting matrix $\gamma^{k\ell}$ then gives the shift in the weight $h$ of the 1/4-BPS state. More precisely, we diagonalize $\gamma^{k\ell}$ to get its eigenvalues $\mu_i$, which give the shift
\be\label{mui}
h^{(2)}_i = \mu_i\ .
\ee
Note that in the notation of \cite{Keller:2019yrr}, $D=-2 M$, the factor of 2 coming from the normalization in (\ref{ccmodulus_real}). In our convention, the matrix $D^{k\ell}$ then needs to be positive semi-definite for 1/4-BPS states to avoid violating the unitarity bound.

\subsection{Lifting 1/4-BPS Primaries Using Ward Identities}
We will now explain how to evaluate (\ref{DijGij}) efficiently in the case where $\varphi$ is a $1/4$-BPS state. The idea is to use Ward identities to reduce the integral (\ref{DijGij}) to a contour integral. This method is very similar to the one introduced in \cite{Gava:2002xb}, and further developed and used in \cite{deBoer:2008ss, Gaberdiel:2015uca,Guo:2019pzk}.

Let the modulus $O$ be given as some $G_{-1/2}$ and $\tilde G_{-1/2}$ descendant of some $h=\bar h =1/2$ 1/2-BPS field $\sigma$. Here we denote right movers by a tilde. Moreover we take the lifted field $\varphi$ to be a 1/4-BPS state. More precisely, we assume that it is a BPS state and a primary field with respect to the right-moving $\cN=4$ SCA, so that it satisfies
\be\label{Gshort}
\tilde G_{-1/2}\varphi = 0\ .
\ee 
It follows that
\be\label{GOPE}
\tilde G(\zb)\varphi(w,\wb) \sim 0\ , \qquad
\tilde G(\zb)\varphi^\dagger (w,\wb) \sim \frac{(\tilde G_{-1/2} \varphi^\dagger)(w,\wb)}{\zb-\wb}\ .
\ee
The second statement follows from the fact that even though $\varphi^\dagger$ is not annihilated by $\tilde G_{-1/2}$, it is still a highest weight state, so that all positive modes of $\tilde G$ annihilate it. The first statement follows from (\ref{Gshort}).

Consider then the correlation function 
\be\label{OOphiphi_herm}
\cC(z_1,z_2,z_3,z_4) =  
\langle \varphi^{\ell\dagger}(z_1,\bar z_1)\; O(z_2,\bar z_2)\; O^\dagger(z_3,\bar z_3)\;\varphi^k(z_4,\bar z_4)\rangle\ .
\ee
Using (\ref{GOPE}), we can use Ward identities as in \cite{Keller:2019suk} to write $\cC$ as a total derivative, namely either as
\bea\label{OOphiphi_iii}
\cC(z_1,z_2,z_3,z_4) =-\partial_{\bar z_3}\Bigg(\frac{\bar z_3-\bar z_1}{\bar z_2-\bar z_1}\;
\Big\langle \varphi^k(z_1,\bar z_1)\;O'(z_2,\bar z_2)\;O^{'\dagger}(z_3,\bar z_3)\;\varphi^{\ell\dagger}(z_4,\bar z_4)\Big\rangle\Bigg)\ .
\eea
or as
\bea\label{OOphiphi_iii2}
\cC(z_1,z_2,z_3,z_4) =-\partial_{\bar z_2}\Bigg(\frac{\bar z_2-\bar z_4}{\bar z_3-\bar z_4}\;
\Big\langle \varphi^k(z_1,\bar z_1)\;O'(z_2,\bar z_2)\;O^{'\dagger}(z_3,\bar z_3)\;\varphi^{\ell\dagger}(z_4,\bar z_4)\Big\rangle\Bigg)\ .
\eea
Here we have introduced the notation
\be\label{half}
O'\equiv G^{(+1)}_{-\frac12}\sigma^{--}
\ee
that is the operator with the same left-moving structure as that of the modulus. 

We can now express $\mathcal G_1^{k\ell}(x,\xb)$ (\ref{curlg1}) in terms of $\cC$  by sending the $z_i$ in (\ref{OOphiphi_iii2}) to their respective positions $z_1\to0$, $z_2\to x$, $z_3\to 1$, and $z_4\to \infty$,
\be\label{I1}
\mathcal G_1^{k\ell}(x,\xb)=\cC(0,x,1,\infty)= -\partial_{\xb}\langle \varphi^\ell|O'^\dagger(1,1)O'(x,\xb)|\varphi^k\rangle=:-\partial_{\xb}  \cI_1^{k\ell}(x,\xb)\ ,
\ee
and $\mathcal G_2^{k\ell}(x,\xb)$ (\ref{curlg2}) by choosing $z_2\to 1,z_3\to x$ in (\ref{OOphiphi_iii}),
\be\label{I2}
{\mathcal G}_2^{k\ell}(x,\xb)=\cC(0,1,x,\infty)= -\partial_\xb\left( \xb \langle \varphi^\ell|O'(1,1)O'^\dagger(x,\xb)|\varphi^k\rangle \right) =: -\partial_\xb \cI_2^{k\ell}(x,\xb)\ ,
\ee
where we defined
\be\label{curlyI1}
\cI_1^{k\ell}(x,\xb)= \langle \varphi^\ell|O'^\dagger(1,1)O'(x,\xb)|\varphi^k\rangle
\ee
and
\be\label{curlyI2}
\cI_2^{k\ell}(x,\xb)=  \xb \langle \varphi^\ell|O'(1,1)O'^\dagger(x,\xb)|\varphi^k\rangle \ .
\ee

\subsection{Contour Integrals}
We use Stokes' theorem in both cases (\ref{I1}) and (\ref{I2}) to reduce the integral to a contour integral around $0,1,\infty$. 
\be
D^{k\ell}=-\frac{i}{2}\int dx d\xb\,\mathcal G^{k\ell}(x,\xb) =\frac{i}{2}\oint_{0,1,\infty} dx (\cI_{1}(x,\xb)+\cI_2(x,\xb)) \ .
\ee
Here the minus sign comes from reversing the direction of the boundary.
We evaluate the contour integral by evaluating the OPE of $O'(x)$ with the three other fields at $0,1,\infty$. 
More precisely, for $x=0$ we parametrize $x=\epsilon e^{i\theta}$, integrate over $\theta$, and discard all non-constant powers in $\epsilon$: they either vanish for $\epsilon\to0$, or, if they diverge, are regulated away.
This means that we pick up the term $x^{-1}\xb^0$ in the expansion of $\cI_{1,2}(x,\xb)$ around $x,\xb=0$. A similar argument shows that we pick up the same term $x^{-1}\xb^0$ in the expansion around $x=\infty$, although with a minus sign, since the contour integral is now clockwise.
Finally, the contour integral around $x=1$ does not contribute, roughly speaking because the pole comes from the OPE of the modulus with itself. For a more detailed argument for this see for instance \cite{Keller:2019suk}.
In total we thus get
\be\label{DI1I2}
D^{k\ell}=-\pi\left( (\cI_{1}(x,\xb)+\cI_2(x,\xb))|_{x^{-1}\xb^0}^0 - (\cI_{1}(x,\xb)+\cI_2(x,\xb))|_{x^{-1}\xb^0}^\infty\right)
\ee
Since we are picking out only a single term in the $x$-expansion of the four point function around 0, effectively we only need to take into account the contribution of a finite number of states of a certain weight. Expanding over an orthonormal basis of states $\chi$, in terms of three point functions we have
\be\label{I1exp0}
\cI_1(x,\xb)|^0= \sum_\chi C_{\varphi^{\ell\dagger} O^{'\dagger}\chi} C_{\chi^\dagger O'\varphi^k}x^{h_\chi-h_k-1}\xb^{\hb_\chi-\hb_k-1/2}
\ee
and
\be\label{I2exp0}
\cI_2(x,\xb)|^0=\sum_\chi C_{\varphi^{\ell\dagger} O^{'}\chi} C_{\chi^\dagger O^{'\dagger}\varphi^k}x^{h_\chi-h_k-1}\xb^{\hb_\chi-\hb_k+1/2}\ .
\ee
The expansion around infinity is very similar. We can use the crossing transformation 
\be
\cI_1(x,\xb)= -\frac1{x^2\xb}\cI_1(1/x,1/\xb)
\ee
to obtain 
\bea\label{I12expinfty}
\cI_1(x,\xb)|^\infty&=&-\sum_\chi C_{\varphi^{\ell\dagger} O^{'}\chi} C_{\chi^\dagger O^{'\dagger}\varphi^k}x^{-h_\chi+h_k-1}\xb^{-\hb_\chi+\hb_k-1/2}\ ,\\
\cI_2(x,\xb)|^\infty&=&-\sum_\chi C_{\varphi^{\ell\dagger} O^{'\dagger}\chi} C_{\chi^\dagger O^{'}\varphi^k}x^{-h_\chi+h_k-1}\xb^{-\hb_\chi+\hb_k+1/2}\ .
\eea
By picking out the appropriate terms, the lifting matrix can thus be written as finite sum of squares of three point functions, 
\begin{multline}\label{Lift3pt}
D^{k\ell}=-2\pi\sum_{\substack{\chi:h_\chi=h_k\\ \hb_\chi=\hb_k+1/2}}
C_{\varphi^{\ell\dagger} O^{'\dagger}\chi} C_{\chi^\dagger O'\varphi^k}
-2\pi \sum_{\substack{\chi:h_\chi=h_k\\ \hb_\chi=\hb_k-1/2}}
C_{\varphi^{\ell\dagger} O^{'}\chi} C_{\chi^\dagger O^{'\dagger}\varphi^k}
\\
=\sum_{\substack{\chi:h_\chi=h_k\\ \hb_\chi=\hb_k+1/2}}
2\pi C^*_{\chi^\dagger O^{'}\varphi^{\ell}} C_{\chi^\dagger O'\varphi^k}
+\sum_{\substack{\chi:h_\chi=h_k\\ \hb_\chi=\hb_k-1/2}}
2\pi C^*_{\chi^\dagger  O^{'\dagger}\varphi^{\ell}} C_{\chi^\dagger O^{'\dagger}\varphi^k}
\end{multline}
The minus sign in the second line comes from the fact that we exchanged an odd number of fermionic operators. In particular we see that the lifting matrix is manifestly positive semidefinite, which is exactly what is needed to preserve unitarity.

\section{Computing the Correlation Functions $\cI_1(x)$ and $\cI_2(x)$}
\label{s:corrcomp}
In this section we discuss our method of computing the 4-point functions $\cI_1(x)$ and $\cI_2(x)$ defined in eqs. (\ref{curlyI1})  and (\ref{curlyI2}), respectively. In subsection \ref{subsec_base}, We start by computing the $N$-dependent combinatorial factors associated with the symmetrization of the states under the permutation group $S_N$, as well as the normalization of the states. We then map the problem to the covering space of the original setup in subsection \ref{subsec_cover}. Finally, in subsection \ref{subsec_4pfcover} we determine bosonisation of fermionic fields and show how to use them to compute the 4-point functions on the cover.

\subsection{Correlation Function on the Base}\label{subsec_base}

Let us now discuss how to compute correlation functions 
\be
\cI_1(x,\xb)= \langle\varphi^{(1)}| O'^{\dagger}(1,1)O'(x,\xb)|\varphi^{(1)}\rangle\ 
\ee
and
\be
\cI_2(x,\xb)= \langle\varphi^{(1)}| O'(1,1)O'^{\dagger}(x,\xb)|\varphi^{(1)}\rangle\\ .
\ee
First we use a conformal transformation to map the half-moduli $O'$ to $0,\infty$ and the $\varphi$ to $1,x$ by using the conformal map 
\be
z \mapsto \frac{x-z}{1-z}
\ee
giving
\be\label{ItocurlyI}
\cI_1(x,\xb) = \frac{(1-x)^{2h_\varphi}(1-\xb)^{2\hb_\varphi}}{(x-1)^2(\xb-1)} I_1(x,\xb)\ ,
\ee
where we defined
\be\label{I1ii}
I_1(x,\xb):=\langle O'|\varphi^{(1)}(1,1)\varphi^{(2)}(x,\xb)|O'\rangle\ .
\ee
Similar expressions hold for $I_2$ and $\cI_2$. The branch cut now runs between $0$ and $\infty$, which will be useful for lifting to the cover.
 
Next we need to take care of the symmetrization of the states. To that end, we want to write $I(x,\xb)$ as a sum over the images of some representative $\check\phi$ over $S_N$. That is, the representative of each of the four operators $O',\varphi^{(1)},\varphi^{(2)}$ and $O'$ is summed over a copy $S^i_N$, $i=1,2,3,4$ of the symmetric group respectively. We want to deal with these sums as follows, picking up various $N$ dependent factors in the process:
\begin{enumerate}
\item We use $S^1_N$ of the first $O'$ as an overall diagonal group to fix the overall `gauge' such that the twisted factors of $O'$ are (12), and all other factors only contain the vacuum. This gives an overall factor of 
\be
N!\ .
\ee 
Note that there are no fermionic signs from these permutations, since we are acting with the diagonal group on all four states at the same time.
\item Next consider the terms coming from $S^4_N$. They all vanish, unless the twisted factors are (12) or (21). In that case, there are an additional $(N-2)!$ possibilities for the vacuum factors, giving an overall factor of 
\be
2(N-2)!\ .
\ee 
There are no fermionic signs since there is only one non-vacuum factor.
\item To count the terms coming from the $S^2_N$ action on $\check\varphi^{(1)}$, we split up the $N!$ terms into configurations $\rho_1$. Such a configuration $\rho_1=(a_1,a_2,S)$ consists of an ordered tuple of factors $(a_1,a_2)$ corresponding to the factors at (1) and (2), and an unordered multiset $S$ corresponding to the remaining $N-2$ factors. Let us call $K_{\rho_1}$ the number of non-vacuum factors in $S$, such that $K_{\rho_1}$ is $L_1,L_1-1$ or $L_1-2$. Note that each configuration comes with a sign, as we pick up fermionic signs for moving factors to (1)(2). The multiplicity of the configuration $\rho_1$ then comes from two contributions:
First, there are $(N-L_1)!/(N-K_{\rho_1}-2)!$ ways of picking the $2-(L_1-K_{\rho_1})$ vacuum factors from the $N-L_1$ vacuum factors in $\check\varphi^{(1)}$. Next, the stabilizer group of (12), $S^2_{N-2}$, gives an additional multiplicity of $(N-2)!$ for each configuration $\rho_1$. We use the stabilizer to fix these $N-2$ factors such that the non-vacuum factors are in the positions $\{3,4,\ldots, K_{\rho_1}+2\}$, not leading to any signs, and pull out the factor of $(N-2)!$. In total we thus pick up a factor of
\be 
\frac{(N-2)!(N-L_1)!}{(N-K_{\rho_1}-2)!}\ .
\ee
\item For $S^3_N$, we again choose configurations $\rho_2$ with corresponding signs. For their multiplicity we again get a factor of $(N-L_2)!/(N-K_{\rho_2}-2)!$. The stabilizer subgroup $S^2_{N-2}$ however needs to be treated differently: the correlation function vanishes unless the non-vacuum factors of $\check\varphi^{(2)}$ in $S$ are lined up with the non-vacuum factors of $\check\varphi^{(1)}$ in $S$, that is unless they are also in the positions $\{3,4,\ldots, K_{\rho_1}+2\}$. In particular this means that $K_{\rho_1}=K_{\rho_2}$. All non-vanishing terms are thus in the sum over the setwise stabilizer of $\{3,4,\ldots, K_{\rho_1}+2\}$, which is $S_K\times S_{N-K-2}$. For a given configuration, we therefore sum over the permutation group $S_K$, and pull out the additional factor of $(N-K-2)!$ from $S_{N-K_{\rho_1}-2}$, giving a total factor of
\be
(N-L_2)!\ .
\ee
\end{enumerate}
Finally, we also need to take into account the normalization of the four operators.
The moduli have $L=2$ and $A=2$. The total normalization factor of the 4-point function is thus
\be\label{normfactor}
(N!)^{-1}(2(N-2)! )^{-1} (||\varphi^{(1)}|| ||\varphi^{(2)}||)^{-1}
\ee
Pulling together all $N$-dependent factors and the normalization  (\ref{normfactor}), for a given configuration $\rho$ with $K=K_\rho$ we get 
\be\label{Ndepfactor} 
\kappa_\rho(N) = 
\frac{(N-2)!(N-L_1)!(N-L_2)!}{||\varphi^{(1)}|| ||\varphi^{(2)}||(N-K_\rho-2)!}\ .
\ee
(Note that since $K_{\rho_1}=K_{\rho_2}$, we simply write $\kappa_\rho$.)
If $\varphi^{(1,2)}$ are of the form (\ref{orbitstate}), then in the large $N$ limit,
\be
\kappa_\rho(N) \sim N^{K-\frac12L_1-\frac12L_2}\ .
\ee
Note that we have $K\leq L_1, L_2$, so that the leading term is $O(1)$ if $K=L_1=L_2$, and any other terms are subleading. Also note that this leading term is a disconnected piece consisting of contracted 2-point functions, which will therefore not contribute to the lifting. This agrees with the general picture in \cite{Belin:2015hwa,Gemunden:2019hie}. To have a connected piece, at least one factor of each $\varphi^{(1)}$ and $\varphi^{(2)}$ has to be in the position 1 or 2, meaning $K<L_1,L_2$. If $L_1=L_2$, this gives a subleading term $O(N^{-1})$. This agrees with the general expectation that the analogue of the 't Hooft coupling should be $\lambda N^{-1/2}$.

In summary, our procedure for computing $I(x)$ is the following:
\be\label{Iexpression}
I(x,\xb) = \sum_{\rho_1,\rho_2} \kappa_{\rho_1}(N) \sum_{g\in S_K}\langle \check O'|\check \varphi^{(1,\rho_1)}(1,1)g \check\varphi^{(2,\rho_2)}(x,\xb)|\check O'\rangle
\ee
Here $\check O'$ is the representative with the twisted factors in (12), and $\check\varphi^{(i,\rho_i)}$ is the representative with $K$ non-vacuum factors in $(3)(4)\ldots(K+2)$, which are permuted by the symmetric group $S_K$.

\subsection{Cover Map}\label{subsec_cover}
Let us now compute the correlator
\be
\langle\check O'|\check \varphi^{(1,\rho_1)}(1,1)g \check\varphi^{(2,\rho_2)}(x,\xb)|\check O'\rangle =: \langle \check O'|\phi^{(1)}(1,1) \phi^{(2)}(x,\xb)|\check O'\rangle
\ee
We can write this as the product of an untwisted two point function $I^u$ involving the factors $(3)(4)\ldots (K+2)$ of $\phi^{(1),(2)}$ with a twisted correlation function $I^t$ in the factors $(12)$,
\be\label{Itw}
I^{tw}(x) = \langle\check O'|(\phi^{(1)1}\otimes \phi^{(1)2})(1,1)
(\phi^{(2)1}\otimes \phi^{(2)2})(x,\xb)|\check O'\rangle
\ee
$I^u$ is of course easy to evaluate as it is a product of two 2-point functions. To compute $I^{tw}$, we go to the double cover using the cover map
\be
z(t) = t^2\ ,
\ee
where $z$ is the coordinate on the base, and $t$ the coordinate on the cover.
This means that we have 
\be
\phi(t) = \phi(z)\left(\frac{dz}{dt} \right)^{h} = \phi(z) (2t)^{h}\ .
\ee
Note that because for fermions we have
\be
\psi(t) = \psi(z)\left(\frac{dz}{dt} \right)^{1/2} = \psi(z) (2t)^{1/2}\ ,
\ee
we are potentially introducing branch cuts.
In particular this means that if we want fermionic operators $\psi(z)$ to be periodic around $0$, then we need to make $\psi(t)$ antiperiodic on the cover. This means that even though we work in the NS sector on the base, we need to work in the Ramond sector on the cover. 

The fields $(\phi^{(1)1}\otimes \phi^{(1)2})(1,1)$ and $(\phi^{(2)1}\otimes \phi^{(2)2})(x,\xb)$ then simply get mapped to
\be\label{onecover}
2^{-h_{\phi^{(1)1}}-\hb_{\phi^{(1)1}}} (-2)^{-h_{\phi^{(1)2}}-	\hb_{\phi^{(1)2}}}\phi^{(1)1}(1,1)\phi^{(1)2}(-1,-1)
\ee
and
\be\label{xcover}
(2\sqrt{x})^{-h_{\phi^{(2)1}}}
(2\sqrt{\xb})^{-\hb_{\phi^{(2)1}}}
(-2\sqrt{x})^{-	\hb_{\phi^{(2)2}}}
(-2\sqrt{\xb})^{-\hb_{\phi^{(2)2}}}
\phi^{(2)1}(\sqrt{x},\sqrt{\xb})\phi^{(2)2}(-\sqrt{x},-\sqrt{\xb})\ ,
\ee
where $h_{\phi^{(i)j}}$ is the holomorphic dimension of the field $\phi^{(i)j}$. When mapping the $\check O'$ we need to be slightly more careful. The twisted ground state $\sigma_2$ usually gets mapped to the vacuum on the cover. As mentioned above, here it gets mapped to a Ramond ground states at $t=0$ and $\infty$.
More precisely, the Ramond ground state is degenerate, forming a doublet under both $su(2)$ R- and flavor symmetry for both left- and right-movers. We will denote these Ramond ground states by 
\be\label{Rgs}
|k_1,k_2;\tilde k_1 ,\tilde k_2\rangle_R\ , \qquad k_i, \tilde k_i\ , =-\frac12,\frac12\ .
\ee
and the corresponding fields by
\be\label{Rfield}
\sigma_R^{k_1k_2\tilde k_1 \tilde k_2} \ , \qquad k_i, \tilde k_i =-\frac12,\frac12\ ,
\ee
or, if we do not want to specify the $su(2)$ charges, simply by $\vac_R$ and $\sigma_R$.

For $G_{-1/2}$ acting on the twisted ground state, using $G(z) = \sum_r G_{r} z^{-r-3/2}$ we write
\begin{multline}
G_{-1/2}|\sigma_2\rangle= G^1_{-1/2}|\sigma_2\rangle+G^2_{-1/2}|\sigma_2\rangle=\oint_0 G^1(z)|\sigma_2\rangle dz+\oint_0 G^2(z)|\sigma_2\rangle  dz\\
= \oint_0 G(t)|0\rangle_R (2t)^{-3/2}2tdt
= \oint_0 G(t)|0\rangle_R (2t)^{-1/2} dt = 2^{-1/2} G_{-1}|0\rangle_R
\end{multline}
The upshot is thus that we have $G_{-1}$ acting on the Ramond ground state. Since $G_{-1} \sim \Xl{I}_{-1} \pl{J}_0$, this means that we simply insert an additional boson at $t=0,\infty$, and we act with a fermion zero mode on the Ramond ground state.
We can check that the normalization is indeed correct by choosing both $\phi$ to be the identity operator. On the base we then have
\be
\langle\sigma^-|G^{(-2)}_{1/2}G^{(+1)}_{-1/2}|\sigma^-\rangle=
\langle\sigma^-|L_0-\frac12J^{(3)}_0|\sigma^-\rangle=1
\ee
and on the cover we have 
\be
\frac12\langle 0_R|G^{(-2)}_1 G^{(+1)}_{-1}|0_R\rangle
= \frac12\langle 0_R|L_0-J^{(3)}_0+\frac34|0_R\rangle=\frac12(\frac14+1+\frac34)=1\ ,
\ee
which indeed agrees. 
Starting from (\ref{Omodulus}) and (\ref{Odaggermodulus}), the  moduli are thus mapped to the following operators on the cover:
\bea
O'(0,0) &\to&\frac1{\sqrt{2}} \left(\Xbl{1}\sigma_R^{\frac12-\frac12-\frac12-\frac12} +\Xbl{2}\sigma_R^{-\frac12\frac12-\frac12-\frac12}\right)(0,0)
\\
O'^\dagger(\infty,\infty) &\to&\frac1{\sqrt{2}} \left(-\Xl{1}\sigma_R^{-\frac{1}2\frac12\frac12\frac12}
+\Xl{2}\sigma_R^{\frac{1}2-\frac12\frac12\frac12}\right)(\infty,\infty)
\eea

\subsection{Correlation Function on the Cover}\label{subsec_4pfcover}
Let us now discuss how to evaluate the correlation function (\ref{Itw}) on the cover. We use the fact that it is a free correlator that can be computed by Wick contractions.
More precisely, we first factor it into a bosonic and fermionic piece. Furthermore we can use the fact that the pairs $\Xl{1},\Xl{3}=\Xbl{1}$ and $\Xl{2},\Xl{4}=\Xbl{2}$ are orthogonal, so that we can also factorize their contributions. For the states $\phi$ that we are considering, $\Xl{I}$ only appears as a $\Xl{I}_{-1}$ descendant, so that we can simply compute a correlation function of free bosons.

The $G$ descendant simply lead to an insertion of $\Xl{I}$ at 0 and $\infty$ and a change in the Ramond ground state. More precisely, the Ramond ground state is degenerate, forming a doublet under both $su(2)$ R- and flavor symmetry for both left- and right-movers. The Ramond ground states and the corresponding fields were defined in eqs. (\ref{Rgs}) and (\ref{Rfield}).

Computing the fermionic part is only slightly more complicated. We again factor it into contributions of the two pairs $\pl{1},\pl{3}$ and $\pl{2},\pl{4}$. The main difference is that there are Ramond ground states at 0 and $\infty$. To compute the Ramond sector correlation functions, we again want to bosonize the fermions. Note that mathematically this is essentially identical to our discussion of bosonization in section~\ref{ss:twist2}; conceptually the difference is simply that there we were working in the NS sector of twisted sector of a symmetric orbifold, whereas now we are working in the R sector of an ordinary fermionic theory.

To compute R sector correlation function,  we again bosonise the fermions as
\be\label{bsnsn_1cf}
\pl{i}= e^{iH^{(i)}}\ ,\qquad\pbl{i}= e^{-iH^{(i)}}\ , \qquad  \pl{i}\pbl{i} = i\partial H^{(i)}\ , \qquad i=1,2\ ,
\ee
where the $H^{(i)}$ are real bosonic fields. One can check that the OPEs between $\Psi$ and $\bar\Psi$ have the correct form.
With this definition, the Ramond ground state operators can be written in the bosonised form 
\be\label{bsnRamond}
\sigma_R^{k_1k_2\tilde k_1 \tilde k_2}=e^{i(k_1H^{(1)}+k_2H^{(2)}+\tilde k_1\tilde H^{(1)} +\tilde k_2 \tilde H^{(2)})}\ .
\ee
These indeed introduce the correct branch cuts as we have
\be\label{bsnsn_S_psi}
\pl{i}(t)\;\sigma_R^-(0)\sim\frac1{t^{\frac12}}\,\sigma_R^+(0)\ ,
\ee
and is anti-periodic when taken around the origin once, as expected. 
The advantage of bosonisation is that all fermionic correlation functions, even those including Ramond spin fields $\sigma_R$, can be evaluated uniformly using free bosons. Concretely, the correlation function of the product of exponentials is of the form, see \cite[appendix 6.A.]{DiFrancesco:1997nk}:
\be\label{exps_ii}
\Big\langle\prod_{l=1}^n:e^{A_l}:\Big\rangle=e^{\sum_{l,m=1,\,l<m}^n\langle A_l\,A_m\rangle}\ .
\ee
For $A=ik\cdot H(t)=i\sum_{i=1}^2 k_iH^{(i)}(t)$, this reads
\be\label{exps_iii}
\Big\langle\prod_{l=1}^n:e^{ik^{(l)}\cdot H(t_l)}:\Big\rangle=\prod_{l,m=1,\,l<m}^n(t_l-t_m)^{k^{(l)}\cdot k^{(m)}}\ ,
\ee
We implemented this algorithm to compute the lifting matrix in the Mathematica notebook \notebook.

\section{Lifting Untwisted Sector States}\label{s:liftUt}
In this section we use the method developed in sections \ref{s:2ndorder} and \ref{s:corrcomp} to evaluate the lifting of a set of 1/4-BPS states in the untwisted sector of the CFT. These states have $h=1$, $j=0$, and $\bar j=0,1,2$. We will first discuss the representation theory and the degeneracy of these states in subsection \ref{N4rep}. We will then compute the lifting of the states with $\bar j=0,1,2$ in subsections \ref{ss:j0lift}, \ref{ss:j1lift}, and \ref{ss:j2lift}, respectively.

\subsection{Contracted Large $\cN=4$ SCA and the 1/4-BPS Spectrum}\label{N4rep}

Let us now apply our computations to the lifting of 1/4-BPS states.
We are interested in 1/4-BPS states of the form 
\be
\chi_{h,j}\tilde \chi_{\bar j} \ ,
\ee
that is states that are in short representations for the right movers, and in long representations for the left movers. Note that because the overall spin is integral, the left moving NS weight $h$ is quantized.
In fact we will concentrate on states with $h=1, j=0$. There are multiple types of right-moving short multiplets that can occur, so that $\bar j$ can in principle take any integer value if $N$ is large enough. As a function of $\bar j$ with fixed $h=1,j=0$, their spectrum can be computed as in \cite{Benjamin:2017rnd}:
\begin{table}[ht]
	\centering
	\begin{tabular}{|cccccccc|}
		\hline
		$\bar j$&0&1&2&3&4&5&$\geq 6$ \\
		\hline
		untwisted&3&6&9&8&3&0&0\\
		twisted&0&8&27&36&23&6&0\\
		total&3&14&36&44&26&6&0\\
		sugra&0& 0& 10& 20& 15& 4& 0\\
		\hline
	\end{tabular}
	\caption{1/4-BPS spectrum for $h=1, j=0$}\label{t:14BPS}
\end{table}
Let us now compute the lifting of the untwisted states for the first few values of $\bar j$, namely $\bar j=0,1$ and 2. For $\bar j=0$ we will present our computations analytically, explaining how we implement the method outlined in section~\ref{s:corrcomp} in practice. For higher values of $j$ we will use the Mathematica notebook \notebook, which we provide as a supplement to this paper. The notebook consists of three parts: The first part defines all the functions necessary to run the code. The second part computes all the primary fields by imposing the primary conditions. In principle this is not necessary, since we could compute the lifting matrix for all states and not just the primaries. However, since the number of primaries is vastly smaller than the number of all states, restricting ourselves to primaries makes the computation much faster. Finally the third part computes the lifting matrix for all primaries and gives the result to all orders in $N$.

The notebook can in principle be used to compute the lifting of any untwisted states in the $T^4$ symmetric orbifold. The only major constraint is that currently only $\pl{I}_{-1/2}$ and $\Xl{I}_{-1}$ descendants are implemented, which is enough for our purposes here. It would be straightforward to allow for higher descendants, but would require to introduce additional bookkeeping to keep track of the descendant modes.

\subsection{$\bar j=0$: The Symmetry Algebra }
\label{ss:j0lift}
Let us start with the case $\bar j=0$, $h=1$. These states correspond to the symmetry algebra, since the right-movers are in the vacuum. For simplicity, we work with the highest weight state of the $\bar j=0$ representation, that is the vacuum.
The representatives $\check \varphi$ of the orbits of states with $\bar j=0, h=1$ are given by
\be
\Xl[1]{I}_{-1}\vac\ , \qquad \pl[1]{I}_{-1/2}\pl[1]{J}_{-1/2}\vac\ ,\qquad
\pl[1]{I}_{-1/2}\pl[2]{J}_{-1/2}\vac\ ,
\ee
of which there are a total of $4+6+6=16$. We recall the notation in eq. (\ref{xpsiI}) where $I=1,2,3,4$. How many of these 16 states are primaries? Clearly $\Xl{I}_{-1}\vac$ are all descendants coming from the generator $\Xl{I}$. Next there are 6 linear combinations coming as $\left(\sum_i \pl[i]{I} \right)\left(\sum_j \pl[j]{J} \right)$ descendants. Finally, there are 3 descendants coming from the $J^{(\pm,3)}$ modes, which are single trace fermionic bilinears. This leaves us with 3 primary fields, as expected. A detailed version of this calculation can be found in \notebook. 
The 1/4-BPS states are thus given by
\be
3\chi_{1,0}\tilde \chi_0\ .
\ee
These are fermionic bilinears which correspond to the generators $\hat J^{(\pm,3)}$ of the flavor $su(2)$. That is, they form a triplet under $su(2)$, and  we expect them to be lifted. In particular we note all three fields have single trace part.

To identify them, we check that they are annihilated by $\pl{I}_{1/2}, \Xl{I}_1$ and $J^{(\pm,3)}_1$. 
For example, we find the following expression
\be\label{JhatminusPrimary}
\hat J^{(-)} = \frac{1}{\sqrt{N!(N-1)!+N!(N-2)!}}\sum_{g\in S_N}\left( \pl[g(1)]{2}_{-1/2}\pl[g(1)]{3}_{-1/2}\vac-  
\pl[g(1)]{2}_{-1/2}\pl[g(2)]{3}_{-1/2}\vac \right)\ .
\ee
First note that the 2-trace term is necessary for $\hat J^{(-)}$ to be a primary field: the single trace term by itself is not annihilated by the large $\cN=4$ generator $\sum_i \pl[i]{2}_{1/2}$.
Second note that we obtained the normalization factor by computing the inner product $\hat J^{(-)}$ with itself.
The first term in the normalization factor comes from the norm of the single trace term. The second term comes from the norm of the 2-trace term. We note that it is subleading in $N$. This holds in general, which means that as long as we are only interested in leading order terms, it is indeed enough to work with the single trace terms only.

Let us now describe the computation of the entry of the lifting matrix coming from the contribution of the single trace terms with itself.  In (\ref{Iexpression}), there are three possible configurations $\rho_{1,2}$ each:
\bea 
(\pl{2}_{-1/2}\pl{3}_{-1/2}\vac,\vac,\{\vac,\ldots,\vac\})&:& K=0\\ (\vac,\pl{2}_{-1/2}\pl{3}_{-1/2}\vac,\{\vac,\ldots,\vac\})&:& K=0\\
(\vac,\vac,\{\pl{2}_{-1/2}\pl{3}_{-1/2}\vac,\ldots,\vac\})&:& K=1\ .
\eea
The two configurations give a vanishing contribution unless $K$ is the same for both. The configuration with $K=1$ is simply an untwisted 2-point function of two $h=1$ fields, 
\be\label{kappaK1}
\frac{((N-1)!)^2(N-2)}{N!(N-1)!+N!(N-2)!}\,\frac1{(x-1)^2}\ .
\ee
For the $K=0$ configurations, we have
\be\label{kappaK0}
\kappa = \frac{((N-1)!)^2}{N!(N-1)!+N!(N-2)!}\ .
\ee
The contributions are of the form
\be\label{Jpm4pf_ii}
\Big\langle O^{'\dagger}(\infty,\infty)\;\;
\pl[i]{4}\pl[i]{1}(1,1)\;\;\pl[j]{3}\pl[j]{2}(x,\bar x)\;\;O^{'}(0,0)\Big\rangle\qquad i,j=1,2\ .
\ee
These get mapped to the cover correlators
\bea\label{Jpm4pf_iii}
&&\bigg\langle \frac1{\sqrt2}\Big(:\Xl{1}e^{-\frac i2(-H^{(1)}+H^{(2)}+\tilde H^{(1)}+\tilde H^{(2)})}:+
:\Xl{2}e^{\frac i2(H^{(1)}-H^{(2)}+\tilde H^{(1)}+\tilde H^{(2)})}:\Big)(t_\infty)\times\nonumber\\
&&\times\;\Big(\frac{\pm1}2\Big)\pl{4}\pl{1}(\pm1)\;\;\;\;\Big(\frac{\pm1}{2\sqrt x}\Big)\pl{3}\pl{2}(\pm\sqrt x)\times\nn\\
&&\times
\frac1{\sqrt2}\Big(:\Xl{3}e^{\frac i2(H^{(1)}-H^{(2)}-\tilde H^{(1)}-\tilde H^{(2)})}:+\Xl{4}e^{\frac i2(-H^{(1)}+H^{(2)}-\tilde H^{(1)}-\tilde H^{(2)})}:\Big)(0)\bigg\rangle
\eea
where the four choices of $\pm$ correspond to the four choices of $i,j=1,2$ (which give a sum of four terms), and $O^{'}$ is defined in eq. (\ref{half}). We will take the limit $t_\infty\to\infty$ at the end of the computation. The term corresponding to $t^+_1=1$ and $t^+_{x}=\sqrt x$ reads
\bea\label{Jpm4pf_iv}
&&\frac1{8\sqrt x}\;\bigg\langle
\Big(:\Xl{1}e^{\frac i2(-H^{(1)}+H^{(2)}+\tilde H^{(1)}+\tilde H^{(2)})}:+:\Xl{2}e^{\frac i2(H^{(1)}-H^{(2)}+\tilde H^{(1)}+\tilde H^{(2)})}:\Big)(t_\infty)\times\nonumber\\
&&\qquad\quad\times\;e^{i(H^{(1)}-H^{(2)})}(1)\;\;\;\;e^{-i(H^{(1)}-H^{(2)})}(\sqrt x)\times\nn\\
&&\qquad\quad\times\Big(:\Xbl{1}e^{\frac i2(H^{(1)}-H^{(2)}-\tilde H^{(1)}-\tilde H^{(2)})}:+:\Xbl{2}e^{\frac i2(-H^{(1)}+H^{(2)}-\tilde H^{(1)}-\tilde H^{(2)})}:\Big)(0)\bigg\rangle=\nonumber\\
&&\frac1{8\sqrt x}\bigg\{\frac{(t_\infty-\sqrt x)}{(t_\infty-1)(t_\infty-0)^2|t_\infty-0|}\bigg(\frac{(1-0)}{(1-\sqrt x)^2}\,\frac1{(\sqrt x-0)}\bigg)+\nonumber\\
&&\qquad+\frac{(t_\infty-1)}{(t_\infty-\sqrt x)(t_\infty-0)^2|t_\infty-0|}\bigg(\frac{1}{(1-\sqrt x)^2(1-0)}\,(\sqrt x-0)\bigg)\bigg\}\
\eea
where in the first three lines we used the bosonised form of the currents --- see eqs. (\ref{Jhatminus}) and (\ref{bsnsn_2compferms}) --- and used the OPEs of exponentials given in eq. (\ref{exps_iii}). Taking the limit $t_\infty\to\infty$ (which requires multiplying the correlation function by the appropriate conformal factor $t_\infty^h \bar t_\infty^{\bar h}$ with $(h,\bar h)$ conformal dimensions of the field at $\infty$), we find that the correlation function (\ref{Jpm4pf_iv}) is of the form
\be\label{Jpm4pf_v}
\frac1{8}\frac{1}{(1-\sqrt x)^2}\,\Big(\frac1 x+1\Big)\ .
\ee
The remaining three configurations in eq. (\ref{Jpm4pf_iii}) are computed similarly. In total we get
\be\label{Jpm4pf_vi}
\frac1{4}\,\bigg(\frac1 x+1\bigg)\bigg(\frac{1}{(1-\sqrt x)^2}+\frac{1}{(1+\sqrt x)^2}\bigg)=\frac1{2x}\,\frac{(1+x)^2}{(1-x)^2}\ .
\ee
Including the combinatorial prefactor (\ref{kappaK0}), we evaluate the 4-point function $I_1$ --- see eq. (\ref{I1ii}):
\be
{I}_1(x,\xb)= 
\frac{((N-1)!)^2}{N!(N-1)!+N!(N-2)!}\,
\frac1{2x}\,\frac{(1+x)^2}{(1-x)^2}\ .
\ee
Finally, including the $K=1$ configuration (\ref{kappaK1}) and the conformal prefactor (\ref{ItocurlyI}) we get
\be
\cI_1(x,\xb) = \frac{1}{(\xb-1)} \frac{((N-1)!)^2}{N!(N-1)!+N!(N-2)!}
\left( \frac1{2x}\,\frac{(1+x)^2}{(1-x)^2}+\frac{N-2}{(x-1)^2}
\right)\ .
\ee
Since the second term in the parentheses comes from the configuration with $K=1$, which is the disconnected piece, it is not surprising that it does not give a contribution to the lifting matrix: when reading off the coefficient of $x^{-1} \xb^0$ in the expansion around 0, only the first term contributes, giving
\be \label{DmmI1}
\frac\pi2\,\frac{((N-1)!)^2}{N!(N-1)!+N!(N-2)!} = \frac\pi2\frac{N-1}{N^2}\ .
\ee
There is no contribution from the expansion around $\infty$. $\cI_2(x)$ is a similar expression as $\cI_1(x)$. Here, there is no contribution from $x=0$, but there is a contribution identical to (\ref{DmmI1}) from $x=\infty$.  The 2-trace terms in (\ref{JhatminusPrimary}) turn out not to give a contribution either: the cross term gives directly vanishing $\cI_{1,2}(x)$, and the term with itself gives an $\cI_{1,2}(x)$ with vanishing expansion coefficients. In total we thus get 
\be \label{Dmm}
D^{--}=\frac{\pi(N-1)}{N^2}\ .
\ee
Let us briefly discuss how we can obtain the same result using (\ref{Lift3pt}). With $h_k=1$ and $\bar h_k=0$, clearly the second term in (\ref{Lift3pt}) vanishes. This corresponds to the observation that $\cI_1$ has no contribution at $x=\infty$, and $\cI_2$ none at $x=0$. To compute the contribution of the first term, we take the decomposition (\ref{Iexpression}) as our starting point: that is, we pull out the combinatorial factor $\kappa$ in (\ref{kappaK0}), and then only compute the three point functions of the representatives. The intermediate field is $\chi=\frac1{\sqrt{2}}\partial X^{(4)}_{-\frac12}\Psi^{(1)}_{0}\sigma_R^{-\frac12-\frac12-\frac12-\frac12}$. The corresponding 3-point function is computed in appendix~\ref{app:3pt}, and turns out to be $C=\frac1{2\sqrt{2}}$, so that $C^2=\frac18$. In total there are 4 configurations with $K=0$, so that the overall contribution indeed agrees with (\ref{Dmm}) once we include the combinatorial factor $\kappa$.

Repeating this computation for $\hat J^{(+)}$, we obtain the same result for $D^{++}$. For $\hat J^{(3)}$ we find
\be
D^{33}=\frac{\pi(N-1)}{N^2}\ .
\ee
In total we find the lifting matrix for the three flavor currents $\hat J^{(+)}, \hat J^{(-)}, \hat J^{(3)}$ to be
\be
\gamma^{kl} = \frac\pi4 \lambda^2 D^{kl}
=\lambda^2\pi^2 \begin{pmatrix}\frac1{4N}&0&0\\
	0&\frac1{4N}&0\\
	0&0&\frac1{4N}
\end{pmatrix}+O(N^{-2})
\ee
As expected, the additional flavor $su(2)$ symmetry that is present at the orbifold point does not survive the perturbation and is broken.

Before moving on to higher weight states, for completeness let us briefly confirm that unlike the flavor currents $\hat J$, the R-currents $J$ do not get lifted.
The computation is of course very similar. The lifting 4-point function for $J^{(-)}$ is now given by
\be
\Big\langle O^{'\dagger}(\infty,\infty)\;\;
\pl[i]{2}\pl[i]{1}(1,1)\;\;\pl[j]{3}\pl[j]{4}(x,\bar x)\;\;O^{'}(0,0)\Big\rangle\qquad i,j=1,2\ ,
\ee
which is mapped to the covering surface as
\bea\label{RJpm4pf_ii}
&&\bigg\langle \frac1{\sqrt2}\Big(:\Xl{1}e^{-\frac i2(-H^{(1)}+H^{(2)}+\tilde H^{(1)}+\tilde H^{(2)})}:+
:\Xl{2}e^{\frac i2(H^{(1)}-H^{(2)}+\tilde H^{(1)}+\tilde H^{(2)})}:\Big)(t_\infty)\times\nonumber\\
&&\times\;\Big(\frac{\pm1}2\Big)\pl{2}\pl{1}(\pm1)\;\;\;\;\Big(\frac{\pm1}{2\sqrt x}\Big)\pl{3}\pl{4}(\pm\sqrt x)\times\nonumber\\
&&\times
\frac1{\sqrt2}\Big(:\Xbl{1}e^{\frac i2(H^{(1)}-H^{(2)}-\tilde H^{(1)}-\tilde H^{(2)})}:+\Xbl{2}e^{\frac i2(-H^{(1)}+H^{(2)}-\tilde H^{(1)}-\tilde H^{(2)})}:\Big)(0)\bigg\rangle
\eea
The term corresponding to $t^+_1=1$ and $t^+_{x}=\sqrt x$ now reads ({\it c.f.} eq. (\ref{Jpm4pf_iv}))
\bea\label{RJpm4pf_iii}
&&\frac1{8\sqrt x}\;\bigg\langle
\Big(:\Xl{1}e^{\frac i2(-H^{(1)}+H^{(2)}+\tilde H^{(1)}+\tilde H^{(2)})}:+:\Xl{2}e^{\frac i2(H^{(1)}-H^{(2)}+\tilde H^{(1)}+\tilde H^{(2)})}:\Big)(t_\infty)\times\nn\\
&&\times\;e^{i(H^{(1)}+H^{(2)})}(1)\;\;\;\;e^{i(-H^{(1)}-H^{(2)})}(\sqrt x)\times\nn\\
&&\times\Big(:\Xbl{1}e^{\frac i2(H^{(1)}-H^{(2)}-\tilde H^{(1)}-\tilde H^{(2)})}:+:\Xbl{2}e^{\frac i2(-H^{(1)}+H^{(2)}-\tilde H^{(1)}-\tilde H^{(2)})}:\Big)(0)\bigg\rangle=\nonumber\\
&&\frac1{8\sqrt x}\bigg\{\frac{(t_\infty-\sqrt x)^0}{(t_\infty-1)^0(t_\infty-0)^2|t_\infty-0|}\bigg(\frac{(1-0)^0}{(1-\sqrt x)^2}\,\frac1{(\sqrt x-0)^0}\bigg)+\nonumber\\
&&\qquad+\frac{(t_\infty-1)^0}{(t_\infty-\sqrt x)^0(t_\infty-0)^2|t_\infty-0|}\bigg(\frac{1}{(1-\sqrt x)^2(1-0)^0}\,(\sqrt x-0)^0\bigg)\bigg\}\ .
\eea
Taking the $t_{\infty}\to\infty$ limit, this term is evaluated to be $(4\sqrt x)^{-1} (1-\sqrt x)^2$. Computing the remaining three terms in eq. (\ref{RJpm4pf_ii}) and putting them all together we find
\be\label{RJpm4pf_iv}
\frac1{4\sqrt x}\bigg(\frac{2}{(1-\sqrt x)^2}-\frac{2}{(1+\sqrt x)^2}\bigg)=\frac{2}{(1-x)^2}\ .
\ee
This is indeed the disconnected part of the partition function. The connected part vanishes, as expected, and hence the R-current is not lifted at the second order.

\subsection{$\bar j=1$}\label{ss:j1lift}
\subsubsection{$\bar j=1$, $h=1/2$: The Moduli}
Next let us consider $\bar j=1$. Before discussing the states with $h=1$, let us make a quick aside on the states with $h=1/2$. Here we will find no 1/4-BPS states: instead there are 16 descendants, and 16 1/2-BPS states.
To see this, note that in total there are 32 states:
\be
\pl[1]{I}_{-1/2}\pr[1]{J}_{-1/2}\vac\ ,\qquad
\pl[1]{I}_{-1/2}\pr[2]{J}_{-1/2}\vac\ ,
\ee
where $\pr{J}$ denotes the right-movers. 16 of them are vacuum descendants coming from the fermionic operators $\pl{I}$ and $\pr{J}$. The remaining 16 states are primaries. They are arranged in 4 1/2-BPS multiplets,
\be
4 \chi_{1}\tilde\chi_{1}\ .
\ee
This means that they correspond precisely to the 16 moduli of the torus, that is the 16 untwisted sector moduli of the symmetric orbifold.
An example of such a primary field is
\be
\check\varphi =(\pl[1]{I}_{-1/2}\pr[1]{I}_{-1/2}+\pl[2]{I}_{-1/2}\pr[2]{I}_{-1/2}
- \pl[1]{I}_{-1/2}\pr[2]{I}_{-1/2}-\pl[2]{I}_{-1/2}\pr[1]{I}_{-1/2})\vac\ ,
\ee
since indeed
\be
(\pl[1]{I}_{1/2}+\pl[2]{I}_{1/2})\check \varphi =(\pr[1]{I}_{-1/2}-\pr[2]{I}_{-1/2}+\pr[2]{I}_{-1/2}-\pr[1]{I}_{-1/2} )\vac=0
\ee
and similarly for the other positive mode generators. Being 1/2-BPS they are of course protected against lifting.

\subsubsection{$\bar j =1$, $h=1$}
Let us now turn to $\bar j =1, h=1$.
Here there are a total of 168 states. They are all $su(2)$ doublets for the right movers, and hence BPS for the right-movers. Analyzing the structure for the left movers reveals that there are 8 new multiplets,
\be
2\chi_{2}\tilde \chi_{1} + 6\chi_{1,0}\tilde \chi_{1}\ ,
\ee
with the remaining states coming from descendants.
The first two multiplets are 1/2-BPS states. The remaining 6 multiplets are 1/4-BPS, which  agrees with table~\ref{t:14BPS}. All 8 states have a single trace part.

We did not compute the lifting matrix by hand, but instead used our Mathematica notebook \notebook. 
The resulting lifting matrix $\gamma$ turns out to be diagonal.
To leading order we have
\be
\gamma^{k\ell} = \lambda^2\pi^2\textrm{diag}\left(\frac1{8N},\frac1{8N},\frac1{8N},\frac1{8N},\frac1{8N},\frac1{8N},0,0\right)+O(N^{-2})\ .
\ee
As expected, the 1/2-BPS states do not get lifted, but all of the 1/4-BPS states do. This agrees with the sugra prediction in table~\ref{t:14BPS}.
The all order result is 
\be
\gamma^{k\ell}= \lambda^2\pi^2\textrm{diag}\left(d(N),d(N),d(N),d(N),d(N),d(N),0,0\right)\ ,
\ee
where
\be
d(N)= \frac{N^2-8}{8N^2(N-1)}\ .
\ee
(Note that our computation assumes $N\geq 3$.) 

\subsection{$\bar j = 2$}\label{ss:j2lift}
Finally let us turn to $\bar j=2$. At $\bar h=1, h = 1$ there are a total of 1068 states. Other than the descendants, they decompose into 57 multiplets as
\be
11 \chi_{2}\tilde \chi_{2} + 9 \chi_{2}\tilde \chi_{1,0}
+ 9 \chi_{1,0}\tilde \chi_{2}+ 28 \chi_{1,0}\tilde \chi_{1,0}
\ee
The first term corresponds to the 1/2-BPS states, and the last term to the non-BPS states. The second and third terms are 1/4-BPS states. We are interested only in the third term with $\bar j=2$, for which there are 9 1/4-BPS states as given in table~\ref{t:14BPS}. In the ordering we are choosing, the first 3 states have a single trace part. The remaining 6 states however have no single trace part, and only have multi-trace contributions.

Let us now discuss the lifting of these 9 1/4-BPS states. Again we used \notebook to do the computation. To order $O(N^{-1})$, the lifting matrix is again diagonal:
\be
\gamma^{k\ell} = \lambda^2\pi^2\textrm{diag}\left(\frac1{8 N},\frac1{8 N},\frac1{8 N},0,0,0,0,0,0\right)+O(N^{-2})
\ee
In the scaling limit that is relevant for the supergravity calculation, this means that only 3 states get lifted. More precisely, only the single trace states get lifted. 
The remaining 6 states do not get lifted. From the supergravity point of view, the multi-trace states can be interpreted as multi particle states of protected particles, whose weight at this order in $N$ is protected, so that they are not lifted. The sugra prediction in table~\ref{t:14BPS} is thus confirmed already on the level of the untwisted states. In particular we expect that of the 27 twisted primaries at $\bar h=1$, 4 do not get lifted, so that a total of 10 1/4-BPS primaries remain unlifted.

We can of course also consider the problem outside of the scaling limit $N\to\infty, \lambda/N\to 1$ by keeping all orders in $N$. In that case the lifting matrix $D$ is no longer diagonal, so that we need to obtain its eigenvalues instead. They are given by
\bea
\mu_1,\mu_2,\mu_3 &=&0\\
\mu_4 &=&\pi(a_1+b_1)= \frac{\pi}{2N}-\frac{5\pi}{2N^2}+ O(N^{-3})\nn\\
\mu_5 &=& \pi(a_1-b_1)=\frac{\pi}{N^2}+ O(N^{-3})\nn\\
\mu_6,\mu_7 &=& \pi(a_2+b_2)= \frac{\pi}{2N}-\frac{5\pi}{2N^2}+ O(N^{-3})\nn\\
\mu_8,\mu_9&=& \pi(a_2-b_2)=\frac{\pi}{N^2}+ O(N^{-3})\nn
\eea
where
\bea
a_1 &=&\frac{N^6-8 N^5+22 N^4-9 N^3-108 N^2+212 N-84}{4 N \left(N^6-5 N^5+6 N^4+7 N^3-23 N^2+18 N-4\right)}\\
b_1 &=&
\frac1{4 N \left(N^6-5 N^5+6 N^4+7 N^3-23 N^2+18 N-4\right)}\Big(N^{12}-24 N^{11}+220 N^{10}+\nonumber\\
&&-890 N^9+868 N^8+5388 N^7-19551 N^6+18504 N^5+26072 N^4-80536 N^3+\nn\\
&&+74256 N^2-24800 N+2832\Big)^{\frac12}\nn\\
a_2&=&  \frac{N^6-8 N^5+23 N^4-16 N^3-100 N^2+236 N-152}{4 (N-1)^2 N \left(N^4-3 N^3+8 N-8\right)}\nn\\
b_2&=&  \frac1{4 (N-1)^2 N \left(N^4-3 N^3+8 N-8\right)}\Big(N^{12}-24 N^{11}+222 N^{10}+\nn\\
&&-928 N^9+1129 N^8+4720 N^7-20120 N^6+25128 N^5+14688 N^4-83520 N^3+\nn\\
&&+107088 N^2-62784 N+14656\Big)^{\frac12}\ .\nn
\eea
We note that $\mu_4$ and $\mu_{6,7}$ start to differ at order $O(N^{-6})$, whereas $\mu_5$ and $\mu_{8,9}$ differ at order $O(N^{-4})$. This result is not surprising from the supergravity point of view: The weight of multi particle states is only protected to leading order in $N$. Interactions between the single particle states will lead to corrections at subleading order, which is exactly what we find here.\footnote{We thank Alex Belin for discussion of the supergravity interpretation of our results.}

\section*{Acknowledgements}
We thank Alex Belin, Chi-Ming Chang, Bin Guo, Ying-Hsuan Lin, Samir Mathur, and Xinan Zhou for very helpful discussions.  We thank Stefano Giusto and Rodolfo Russo for helpful comments on the draft. 
C.A.K. thanks Luis Apolo, Suzanne Bintanja, Alex Belin, and Alejandra Castro for discussions and collaboration on a related project.
The work of N.B. is supported in part by the Simons Foundation Grant No. 488653. 
The work of C.A.K. is supported in part by the Simons Foundation Grant No.~629215.

\appendix

\section{1/4-BPS States in Sym$^2(T^4)$}  \label{app:quartBPS}

In this section we will use representation theory arguments to calculate the quarter-BPS spectrum of a generic point in the moduli space of Sym$^2(T^4)$. Before doing so, we will first review the argument computing the generic quarter-BPS spectrum for a $K3$ surface, first done in \cite{Ooguri:1989fd}.

\subsection{Generic BPS spectrum in $K3$}
\label{sec:k3arg}

The $K3$ sigma model is governed by the small $\mathcal{N}=4$ superconformal algebra at $c=6$. This algebra has two massless representations whose characters were computed in \cite{Eguchi:1987sm, Eguchi:1987wf}. The two massless characters combine to form a massive character via
\be
\chi_G(q,y) + 2\chi_M(q,y) = \chi^{\text{long}}(q,y).
\label{eq:n4pairup}
\ee
The elliptic genus is insensitive to times when the quarter-BPS states ``pair up," as in (\ref{eq:n4pairup}). However, of the two massless representations, one is the vacuum multiplet. If we assume that a generic point in the conformal manifold, the chiral algebra is not enhanced, then this fully determines the quarter-BPS spectrum. This was used in \cite{Ooguri:1989fd} to calculate the generic BPS spectrum. The generic BPS spectrum
\be
Z^{\text{BPS}} = \chi^G \overline{\chi^G} + 20 \chi^M \overline{\chi^M} + \sum_{h=1}^{\infty} N_h q^h \(\chi^G + 2\chi^M\)\overline{\chi^M} + \text{c.c.} 
\ee
has elliptic genus
\be
Z^{\text{EG}} = -2\chi^G + 20 \chi^M + \sum_{h=1}^{\infty} N_h q^h \(\chi^G + 2\chi^M\).
\ee
Therefore
\begin{align}
\sum_{h=1}^{\infty} N_h q^h &= \frac{Z^{\text{EG}}(q,y) + 2\chi^G(q,y) - 20\chi^M(q,y)}{\chi^G(q,y) + 2\chi^M(q,y)} \nn\\
&= 90q + 462q^2 + 1540q^3 + \ldots.
\label{eq:90462}
\end{align}
Note that although naively the second line of (\ref{eq:90462}) is $y$-dependent, the final answer is not. It was also pointed out in \cite{Ooguri:1989fd} that a necessary consistency condition for (\ref{eq:90462}) to be the generic quarter-BPS spectrum is for all the coefficients to be non-negative integers. We also pause to point out that these integers have interesting relations to irreducible representations of the sporadic Mathieu group $M_{24}$ \cite{Eguchi:2010ej}.

The fact that a generic point in moduli space has no additional currents was checked in conformal perturbation theory away from the orbifold point $T^4/\mathbb Z_2$ \cite{Keller:2019suk}. There, the first term of (\ref{eq:90462}) was checked. 

\subsection{Generic BPS spectrum in Sym$^2(T^4)$}

The chiral algebra of a generic point in the symmetric product of $T^4$ is larger than that of $K3$. Instead of the small $\mathcal{N}=4$ algebra, the theory has contracted large $\mathcal{N}=4$ algebra. The characters of this algebra were computed in \cite{Petersen:1989zz, Petersen:1989pp}. At $c=12$, the same phenomena happens as in $K3$: there are only two massless representations, of which one is the vacuum multiplet. Therefore if we assume that a generic point in moduli space has no additional conserved currents, we can repeat the argument in section \ref{sec:k3arg} to compute the generic quarter-BPS spectrum. If we call the massless characters $\chi^0, \chi^1$ with $\chi^0$ being the vacuum, they combine into a long multiplet as:
\be
\chi^0(q,y) + 2\chi^1(q,y) = \chi^{\text{long}}(q,y).
\label{eq:t4combine}
\ee
Note that because of the fermion zero modes, the quantity that is protected under (\ref{eq:t4combine}) is no longer the elliptic genus (which trivially vanishes), but the ``modified" index of \cite{Maldacena:1999bp}. The generic BPS spectrum is given by
\be
Z^{\text{BPS}} = \chi^0 \overline{\chi^0} + 5 \chi^1 \overline{\chi^1} + \sum_{h=1}^{\infty} N_h q^h \(\chi^0 + 2\chi^1\)\overline{\chi^1} + \text{c.c.}
\ee
where the first two lines are determined by the half-BPS spectrum (which can be read off from the hodge diamond). This has modified index
\be
Z^{\text{modified EG}} = 2\chi^0 - 5\chi^1 -  \sum_{h=1}^{\infty} N_h q^h \(\chi^0 + 2\chi^1\)
\ee
so therefore
\begin{align}
\sum_{h=0}^{\infty} N_h q^h &= \frac{Z^{\text{modified EG}}(q,y) - 2 \chi^0(q,y) + 5\chi^1(q,y)}{-\chi^0(q,y) - 2\chi^1(q,y)} \nn\\
&= 42q^2 + 70q^3 + 324q^4 + 672q^5 + 1820q^6 + 3726q^7 + 8370q^8 + 16380 q^9 + \ldots.
\label{eq:4270}
\end{align}
In \cite{Guo:2019ady, Guo:2020gxm}, the first four terms (\emph{i.e.} the terms up to $q^4$) of (\ref{eq:4270}) were computed from conformal perturbation theory. 

For reference recall $Z^{\text{modified EG}}(q,y)$ was computed in (5.8) of \cite{Maldacena:1999bp} and has the first few terms for Sym$^2(T^4)$:
\bea
&&Z^{\text{modified EG}}(q,y) = \(2y^2 + y - 6 + y^{-1} + 2y^{-2}\) +\\
&&\qquad\qquad\qquad\quad\;+\(y^{-3} - 12y^{-2} + 39y^{-1} - 56 + 39y - 12y^2 + y^3\) q + \ldots\ .\nn
\eea

\section{3-point Functions in Sym$^N(T^4)$}\label{app:3pt}
In this appendix we consider the lifting 4-point function of the flavor current $\hat J^{(-)}$ which was evaluated in section \ref{ss:j0lift} explicitly. An alternative derivation of the result is given in terms of a finite sum of squares of 3-point function using formula (\ref{Lift3pt}) --- see the discussion below eq. (\ref{Dmm}).

We shall compute $\cI_1$ defined in eq. (\ref{curlyI1}): 
\bea\label{J4pf_ii}
&&{\cal I}_1=\langle\hat J^{(+)}(z_1,\bar z_1)\;O^{'\dagger}(z_2,\bar z_2)\;O^{'}(z_3,\bar z_3)\;\hat J^{(-)}(z_4,\bar z_4)\rangle\\
&&\quad\;=\sum_{\chi}\langle\hat J^{\pm,3}(z_1,\bar z_1)\;O^{\prime\dagger}(z_2,\bar z_2)|\chi\rangle\langle\chi|O^{\prime}(z_3,\bar z_3)\;\hat J^{\pm,3}(z_4,\bar z_4)\rangle\nn\ .
\eea
The 3-point functions we are interested in computing are of the form
\bea\label{J3pf}
&&{ C}:=\langle\hat J^{(-)}(z_1,\bar z_1)\;O^{\prime\dagger}(z_2,\bar z_2)\chi(0,0)\rangle\\
&&\quad\;=\Big\langle\pbl{1}_{-\frac12}\pl{2}_{-\frac12}(z_1,\bar z_1)\;(\partial{X}^1_{-\frac12}\bar\Psi^1_0+\partial{X}^2_{-\frac12}\bar\Psi^2_0)\sigma^{\frac12\frac12\frac12\frac12}(z_2,\bar z_2)\;\chi(0,0)\Big\rangle\ .\nonumber
\eea
The intermediate fields $\chi$ which contribute to lifting have conformal dimensions $h_\chi=1$, $\bar h_\chi=\tfrac12$. Moreover, group selection rule requires $\chi$ to be in the twisted sector. As such, the anti-holomorphic part of $\chi$ can only be in the twist ground state. The holomorphic part has excitations which are restricted by the bosonic and fermionic excitations of $\hat J^{(-)}$ and $O^\dagger$. It turns out the only intermediate field in this case is: $\chi=\frac{1}{\sqrt2}\partial X^{(4)}_{-\frac12}\Psi^1_{0}\sigma^{-\frac12-\frac12-\frac12-\frac12}$:
\be\label{J3pf_i}
{C}=\Big\langle\pbl{1}_{-\frac12}\pl{2}_{-\frac12}(z_1,\bar z_1)\;\;(\partial{X}^1_{-\frac12}\bar\Psi^1_0+\partial{X}^2_{-\frac12}\bar\Psi^2_0)\sigma^{\frac12\frac12\frac12\frac12}(z_2,z_2)\;\;
\frac1{\sqrt2}\partial\bar X^2_{-\frac12}\Psi^1_{0}\sigma^{-\frac12-\frac12-\frac12-\frac12}(0,0)\Big\rangle\ .
\ee
Bosonising the fermions and the current --- see eqs. (\ref{bsnsn_2compferms}) and (\ref{Jhatminus}) --- we find:
\bea\label{OOvarphivarphi_iii}
&&C=\Big\langle \no{e^{i(-H^{(1)}+H^{(2)})}}(z_1,\bar z_1)\;
\no{\big(e^{\frac i2(-H^{(1)}+H^{(2)})}{ X}^1_{-\frac12}+e^{\frac i2(H^{(1)}-H^{(2)})}\partial{ X}^2_{-\frac12}\big)\sigma_b}(z_2,\bar z_2)\times\nonumber\\
&&\qquad\qquad\qquad\qquad\qquad\qquad\qquad\qquad\qquad\qquad\times\frac1{\sqrt2}
\no{e^{\frac i2(H^{(1)}-H^{(2)})}\partial\bar{ X}_{-\frac12}^2 \sigma_b}(0,0)\Big\rangle\nonumber\\
&&\quad=\frac{1}{2\sqrt2}\frac{1}{z_2^{\frac32}\,\bar z_2}\Big\langle :e^{i(-H^{(1)}+H^{(2)})}:(z_1,\bar z_1):e^{\frac i2(H^{(1)}-H^{(2)})}:(z_2,\bar z_2):e^{\frac i2(H^{(1)}-H^{(2)})}:(0,0)\Big\rangle\nonumber\\
&&\quad=\frac{1}{2\sqrt2}\frac{1}{z_2^{\frac32}\,\bar z_2}\,\frac{z_2^{\frac12}}{(z_1-z_2)z_1}\ .
\eea
Next, sending $z_1\to\infty$ and $z_2\to1$, we find that ${C}=\frac1{2\sqrt2}$. This establishes the claim below eq. (\ref{Dmm}).

\bibliographystyle{ytphys}
\bibliography{refmain}

\end{document}